\documentclass[aps,prd,superscriptaddress]{revtex4-2}
\usepackage{amsmath,amssymb,amsthm,bm,graphicx,subfigure,xcolor}
\usepackage[colorlinks=true,linkcolor=blue,citecolor=blue,urlcolor=black]{hyperref}

\begin{document}
\title{Nonlinear dynamics of hot, cold and bald Einstein-Maxwell-scalar black holes in AdS spacetime}

\author{Qian Chen}
\email{chenqian192@mails.ucas.ac.cn}
\affiliation{School of Physical Sciences, University of Chinese Academy of Sciences, Beijing 100049, China}

\author{Zhuan Ning}
\email{ningzhuan17@mails.ucas.ac.cn}
\affiliation{School of Physical Sciences, University of Chinese Academy of Sciences, Beijing 100049, China}

\author{Yu Tian}
\email{ytian@ucas.ac.cn}
\affiliation{School of Physical Sciences, University of Chinese Academy of Sciences, Beijing 100049, China}
\affiliation{Institute of Theoretical Physics, Chinese Academy of Sciences, Beijing 100190, China}

\author{Bin Wang}
\email{wang\_b@sjtu.edu.cn}
\affiliation{Center for Gravitation and Cosmology, College of Physical Science and Technology, Yangzhou University, Yangzhou 225009, China}
\affiliation{Shanghai Frontier Science Center for Gravitational Wave Physics, Shanghai Jiao Tong University, Shanghai 200240, China}

\author{Cheng-Yong Zhang}
\email{zhangcy@email.jnu.edu.cn}
\affiliation{Department of Physics and Siyuan Laboratory, Jinan University, Guangzhou 510632, China}

\begin{abstract}
We investigate the dynamical transition processes of an Einstein-Maxwell-scalar gravitational system between two local ground states and an excited state in the anti-de Sitter spacetime.
From the linear perturbation theory, only the excited state possesses a single unstable mode, indicating the dynamical instability.
Such an instability is associated with the tachyonic instability due to the presence of an effective potential well near the event horizon.
From the nonlinear dynamics simulation, through the scalar field accretion mechanism, the critical phenomena in the transition process of the gravitational system between the two local ground states are revealed.
The threshold of the accretion strength indicates the existence of a dynamical barrier in this transition process, which depends on the coupling strength between the scalar and Maxwell fields.
On the other hand, for the unstable excited state, there exists a special kind of critical dynamics with a zero threshold for the perturbation strength.
The perturbations of different signs push the gravitational system to fall into different local ground states.
Interestingly, in an extended parameter space, there exist specific parameters such that the perturbations of non-zero amplitude fail to trigger the single unstable mode of the excited state.
\end{abstract}

\maketitle

%=======================================================================
\section{Introduction}\label{sec:I}
The no-hair theorem is crucial to characterize the properties of steady-state black holes \cite{Bekenstein:1972ny,Mayo:1996mv}, but the conditions under which it holds have been controversial. %\cite{Hong:2019mcj,Hong:2020miv}.
In particular, the recent discovery of a series of hairy black holes has refreshed the understanding of classical black holes \cite{Herdeiro:2015waa}. 
Among them, one of the most famous mechanisms that leads to a hairy black hole is the spontaneous scalarization due to the non-minimal coupling between the scalar field and the spacetime curvature \cite{Doneva:2017bvd,Silva:2017uqg,Antoniou:2017acq,Cunha:2019dwb,Dima:2020yac,Herdeiro:2020wei,Berti:2020kgk} or the  matter source \cite{Cardoso:2013fwa,Cardoso:2013opa,Damour:1993hw,Zhang:2014kna,Herdeiro:2018wub} which gives the scalar field an effective tachyonic mass. 
The tachyonic instability triggers a strong gravity phase transition and results in a hairy black hole. 
The spontaneous scalarization can significantly affect the properties of compact objects while passing the weak field tests, and thus has attracted much attention recently \cite{Fernandes:2019rez,Fernandes:2019kmh,Oliveira:2020dru,Brihaye:2019puo,Brihaye:2019gla,Peng:2019qrl,Macedo:2019sem,Guo:2020sdu,Astefanesei:2019pfq,Lin:2020asf,Liu:2020yqa,Guo:2020zqm}, especially for the Einstein-scalar-Gauss-Bonnet (EsGB) theory. 
Whether the hairy black hole is energetically favorable over the general relativity solution depends on the coupling function and the range of parameters in the theory. 
The linear stability of the hairy black hole has been discussed in many works \cite{Blazquez-Salcedo:2018jnn,Blazquez-Salcedo:2020caw,Blazquez-Salcedo:2020rhf,Silva:2018qhn,Zou:2020zxq,Myung:2019oua,Myung:2018vug,Zhang:2020pko}. 

To disclose the details of the nonlinear dynamics of spontaneous scalarization, the fully nonlinear evolution have been done in the EsGB theory. 
Moreover, the imprint of the scalar hair in the gravitational radiation and the dynamical descalarization were studied in the binary black hole mergers \cite{East:2020hgw,East:2021bqk,Silva:2020omi}.
For its sibling theory, the Einstein-dilaton-Gauss-Bonnet (EdGB) theory,
the study of the gravitational collapse found tentative evidence that black holes endowed with scalar hair form \cite{Ripley:2019irj}. 
The nonlinear evolution of black holes under perturbations also shows that a hairy black hole forms indeed as the end state of scalarization \cite{Ripley:2019aqj,Ripley:2020vpk,Liu:2022eri}.
The equations of motion may not be well posed in the EsGB or EdGB theory such that these works are limited in small parameter regions of these theories. 
In the Einstein-Maxwell-scalar (EMs) theory, the equations of motion are always well posed and allow the nonlinear study for large couplings. 
The nonlinear dynamics of single black hole in the EMs theory shows that a hairy black hole forms at the end \cite{Herdeiro:2018wub,Fernandes:2019rez,Fernandes:2019kmh,Zhang:2021edm,Zhang:2021ybj,Zhang:2021etr,Xiong:2022ozw}.
Differences with general relativity in the binary black hole merger are significant only for large charge \cite{Hirschmann:2017psw}. 

In addition to the above tachyonic instability that provides the dynamical mechanism of scalarizing black holes, the nonlinear instability of black holes has also been discovered recently \cite{Zhang:2021nnn,Zhang:2022cmu,Liu:2022fxy,Jiang:2023yyn}, suggesting a new scalarization mechanism.
Whether the source of the scalar field is provided by the spacetime curvature in the EsGB theory or the electromagnetic field in the EMs theory, a class of critical scalarization phenomena occurs through the nonlinear accretion of scalar field into a central black hole if the coupling function in the scalar source is dominated by a quartic term.
In the dynamical intermediate process, an unstable critical state that acts as a dynamical barrier emerges, separating the final bald and hairy black holes. \footnote{It is argued \cite{Li:2020ayr} in the context of AdS/CFT under the probe limit that there should be at least one unstable excited state if there are two local ground states that are connected in the configuration space, and the unstable excited state acts as the ``lowest'' dynamical barrier for the transition between the two local ground states. This argument is believed to hold generally in other (gravitational or non-gravitational) systems as well \cite{Chen:2022cwi}.}
For the EMs theory in asymptotically flat spacetime, the finally stable and intermediately unstable hairy black holes in such critical scalarization process are expected to be exactly the hot and cold scalarized black holes found in \cite{LuisBlazquez-Salcedo:2020rqp,Blazquez-Salcedo:2020nhs}.

However, in the asymptotically anti-de Sitter (AdS) spacetime, the complete phase diagram structure of the case where the coupling function is dominated by a quartic term in the EMs theory is still unknown, although the related dynamical process indicates that there should be two branches of hairy black holes.
From the AdS/CFT point of view, the answer to this question will be of great significance for probing the properties of the holographic QCD phase diagram \cite{DeWolfe:2010he,DeWolfe:2011ts,Cai:2012xh,Grefa:2021qvt,He:2013qq,Cai:2022omk}.
On the other hand, although it can be deduced from the critical phenomenon that the intermediate critical state has only one unstable mode, it is still necessary to give direct evidence and reveal the quasi-normal modes, which is crucial for characterizing the dynamical properties.
Most importantly, at present, by fine-tuning a single parameter of the initial value, the critical state can only briefly appear in the middle of the dynamical process, and the nonlinear evolution with it as the initial configuration is still pending.
Since this model allows for two stable local ground states (the hot hairy and bald black holes), the final fate of the unstable cold hairy black holes cannot be predicted until nonlinear evolution is implemented.
These questions have motivated us to investigate further.

In this paper, we study the real-time dynamics on the local ground and excited states in spherically symmetric AdS spacetime in the EMs theory, which contains a non-minimal coupling function between the scalar and Maxwell fields dominated by a quartic term.
By numerically solving the static field equations, the phase structure of the model is revealed, which shows that the domain of existence for black hole solutions consists of two branches of hairy black holes and one branch of RN-AdS black holes.
From the linear perturbation theory, one branch of hairy black holes (hot hairy black holes) is linearly stable like RN-AdS black holes, while the other branch of hairy black holes (cold hairy black holes) with a single unstable mode is linearly dynamically unstable.
For a gravitational system with fixed energy, the hot hairy and RN-AdS black holes serve as two local ground states and the cold hairy black hole acts as an excited state.
Through fully nonlinear numerical simulations, the real-time dynamics based on these states are revealed, where critical phenomena emerge.
For the case where the initial value is a stable local ground state, we find that a scalar field accretion process of sufficient strength can induce a dynamical transition from one local ground state to the other.
The occurrence of such a transition requires the gravitational system overcome a dynamical barrier, leading to the existence of a threshold for the accretion strength.
Near the threshold, the system is excited to an unstable excited state.
For the case where the initial value is an unstable excited state, on the other hand, the system will fall into one of the two local ground states under arbitrarily small perturbations.
The selection of the final state of the evolution depends on the specific form of the perturbation.
Scanning the parameter space of the perturbation, the two local ground states occupy different regions in the spectrum of the final state with the excited state as the boundary.

The organization of the paper is as follows. In section \ref{sec:M}, we give a brief introduction to the EMs model. 
In section \ref{sec:SS}, we numerically solve the static solutions of the field equations to obtain the phase diagram structure of the model.	
In section \ref{sec:S}, we reveal the effective potentials and quasinormal mode spectrums of the three classes of thermal phases.
In section \ref{sec:CD}, we study the dynamical transition process of a gravitational system from one of the two local ground states to the other by crossing an excited state.
In section \ref{sec:BD}, the real-time dynamics during the transition from an excited state to a local ground state is further revealed.
Finally, we conclude the paper in section \ref{sec:Co}.

%=======================================================================

%=======================================================================
\section{The EMs model}\label{sec:M}
We consider 4-dimensional EMs gravity with a negative cosmological constant described by the action
\begin{equation}
	S=\frac{1}{2\kappa^{2}_{4}}\int d^{4}x\sqrt{-g}\left[R-2\Lambda-\frac{1}{4}f(\phi)F_{\mu\nu}F^{\mu\nu}-\nabla_{\mu}\phi\nabla^{\mu}\phi-m^{2}\phi^{2}\right],\label{eq:action}
\end{equation}
where $R, F_{\mu\nu}, \phi$ are the Ricci scalar curvature, Maxwell field strength tensor and a real scalar field, respectively.
For convenience, the cosmological constant is set to be $\Lambda=-3$ to work in units of AdS radius.
In what follows, we shall take the mass squared of the scalar field $m^{2}=-2$ to respect the Breitenlohner-Freedman (BF) bound \cite{Breitenlohner:1982jf} for definiteness.
The interaction between the real scalar field and the electromagnetic field is governed by a non-minimal coupling function $f(\phi)$.

In this model, the variation of the action (\ref{eq:action}) with respect to the metric tensor $g_{\mu\nu}$ gives rise to the following Einstein equation
\begin{equation}
	R_{\mu\nu}-\frac{1}{2}Rg_{\mu\nu}=-\Lambda g_{\mu\nu}+T^{M}_{\mu\nu}+T^{\phi}_{\mu\nu},\label{eq:2.2}
\end{equation}
where the stress-energy tensors of the Maxwell and scalar fields have the form
\begin{subequations}
	\begin{align}
		T^{M}_{\mu\nu}&=\left(\frac{1}{2}g^{\rho\sigma}F_{\mu\rho}F_{\nu\sigma}-\frac{1}{8}F_{\alpha\beta}F^{\alpha\beta}g_{\mu\nu}\right)f(\phi),\\
		T^{\phi}_{\mu\nu}&=\nabla_{\mu}\phi\nabla_{\nu}\phi-\frac{1}{2}\left(\nabla_{\alpha}\phi \nabla^{\alpha}\phi+m^{2}\phi^{2}\right)g_{\mu\nu}.
	\end{align}
\end{subequations}
On the other hand, the equations of motion for the Maxwell and scalar fields can be obtained by varying the action (\ref{eq:action}) with respect to the corresponding matter fields, respectively, as follows
\begin{subequations}
	\begin{align}
		\nabla_{\nu}\left[f(\phi)F^{\nu\mu}\right]&=0,\label{eq:M_E}\\
		\nabla^{\mu}\nabla_{\mu}\phi&=\frac{1}{8}\frac{df(\phi)}{d\phi}F_{\mu\nu}F^{\mu\nu}+m^{2}\phi.\label{eq:S_E}
	\end{align}
\end{subequations}

In this paper, we require the model to contain a branch of stable RN-AdS black hole solutions and be $\mathbb Z_{2}$-invariant under the transformation $\phi\rightarrow-\phi$.
To this end, the coupling function between the scalar and Maxwell fields is assumed to be an exponential dependence without loss of generality
\begin{equation}
	f(\phi)=e^{\alpha\phi^{4}},\label{eq:coupling}
\end{equation}
with a positive coupling constant $\alpha$, which has a global minimum for $\phi=0$.
Such a choice is only for simplicity and convenience of numerical calculation due to the existence of term $\frac{df}{fd\phi}$ in the field equaitons.
A simpler form, such as $f(\phi)=1+\alpha\phi^{4}$, does not qualitatively change the conclusion of the paper.

In order to solve above time-dependent field equations numerically with the characteristic formulation \cite{Chesler:2013lia}, which has been widely used to study the non-equilibrium dynamics of black holes \cite{Chesler:2008hg,Chesler:2010bi,Adams:2013vsa,Chesler:2015bba,Janik:2017ykj,Chesler:2018txn,Attems:2018gou,Bea:2020ees,Chen:2022cwi,Chen:2022vag,Chen:2022tfy}, the ingoing Eddington-Finkelstein metric ansatz compatible with spherical symmetry is adopted
\begin{equation}
	ds^{2}=-2W(t,r)dt^{2}+2dtdr+\Sigma(t,r)^{2}d\Omega^{2}_{2},\label{eq:ansatz}
\end{equation}
where $d\Omega^{2}_{2}$ represents the line element of a unit radius $S^{2}$.
Such form of the metric ansatz is invariant to the following shift transformations in the radial coordinate
\begin{subequations}
	\begin{align}
		r&\rightarrow r+\lambda(t),\\
		W&\rightarrow W+d_{t}\lambda(t),\\
		\Sigma&\rightarrow\Sigma,
	\end{align}\label{eq:shift}
\end{subequations}
which allow us to fix the radial position of the apparent horizon during the dynamics.
For the Maxwell field, we take the gauge $A_{\mu}dx^{\mu}=A(t,r)dt$.

By taking a Taylor expansion of the field equations near the AdS boundary, the asymptotic behaviors of the field variables can be obtained as follows
\begin{subequations}
	\begin{align}
		\phi&=\phi_{1}r^{-1}+\phi_{2}r^{-2}+o(r^{-3}),\\
		A&=\mu-Qr^{-1}+o(r^{-2}),\\
		\Sigma&=r+\lambda-\frac{1}{4}\phi^{2}_{1}r^{-1}+o(r^{-2}),\\
		W&=\frac{1}{2}(r+\lambda)^{2}+\frac{1}{2}-\frac{1}{4}\phi^{2}_{1}-d_{t}\lambda-Mr^{-1}+o(r^{-2}),
	\end{align}\label{eq:BC}
\end{subequations}
where the constants $Q$ and $M$ are electric charge and Arnowitt-Deser-Misner mass \cite{Abbott:1981ff}, respectively.
Since we only focus on the properties of the system in the microcanonical ensemble, without loss of generality, the electric charge $Q$ is fixed to be $1$ in what follows unless otherwise stated.
According to the holographic dictionary \cite{Klebanov:1999tb,Bianchi:2001kw}, the free parameter $\phi_{1}$ is the source of the scalar field on the AdS boundary, which is set to zero to work with the source free boundary condition.
At this point, the response of the scalar field $\phi_{2}$, whose value can only be determined after solving the bulk, is proportional to the expectation value of the scalar operator of the boundary conformal field theory.
The parameter $\mu$, which is set to zero in our work, is a pure gauge whose difference from the value of Maxwell field at the horizon $A(r_{h})$ is the chemical potential.
Since the effective Newton constant is chosen as $\kappa^{2}_{4}=1$ for convenience, the energy density and entropy density of the system are denoted as
\begin{equation}
	\epsilon=2M,\qquad s=2\pi\Sigma^{2}(r_{h}),
\end{equation}
respectively, where $r_{h}$ stands for the radius of apparent horizon.
On the other hand, in order to describe the quasi-local mass, we introduce the rescaled Misner-Sharp (MS) mass \cite{Misner:1964je,Maeda:2012fr} defined as 
\begin{equation}
	M_{MS}=\frac{1}{2}\Sigma\left(-\frac{1}{3}\Lambda\Sigma^{2}+1-g^{\mu\nu}\partial_{\mu}\Sigma\partial_{\nu}\Sigma\right),
\end{equation}
which tends to the ADM mass on the AdS boundary.
For a static solution, the temperature is extracted as 
\begin{equation}
	T=\frac{1}{2\pi}d_{r}W(r_{h}).
\end{equation}
With these boundary conditions in hand, the field equations can be easily solved numerically.
For more details on the numerical procedures for the static solutions and dynamical evolution, we refer readers to subsection \ref{sec:NP} and Appendix A in \cite{Chen:2022vag}, respectively.

\section{Static solutions}\label{sec:SS}
In this section, we reveal the complete phase structure of this model with the coupling function (\ref{eq:coupling}) by numerically solving the static field equations.
The results show that the domain of existence of solutions is composed of three branches: hot hairy black holes, cold hairy black holes and RN-AdS black holes.

\subsection{Numerical procedure}\label{sec:NP}
In order to describe the relationship between the physical quantities of the equilibrium phases, we need to seek out the static solutions to the field equations.
By eliminating the time dependence in the field equations, the static field configuration $\textbf{X}=\{\Sigma,W,A,\phi\}$ is determined by the following independent ordinary differential equations
\begin{subequations}
	\begin{align}
		0&=2\frac{\Sigma''}{\Sigma} + \phi'^{2} ,\\
		0&=W'' + 2 W' \frac{\Sigma'}{\Sigma} -\frac{1}{4}A'^{2} f - \phi^{2} - 3 ,\\
		0&=\frac{1}{2}A'' + A' \left( \frac{\Sigma'}{\Sigma}  + \phi\phi'\frac{d\text{ln}f}{d\phi^{2}}\right),\\
		0&=\left(W \phi'\right)'+ 2 W \phi'\frac{\Sigma'}{\Sigma}  + \left(\frac{1}{4}A'^{2} \frac{df}{d\phi^{2}}  + 1\right)\phi ,
		%		0&=2\left(W\Sigma'\Sigma\right)'  + \left[\left( B   W\right)'^{2}f  - \phi^{2} - 3\right]\Sigma^{2} - 1,
	\end{align}
\end{subequations}
where prime stands for the derivative with respect to the radius $r$.
The above system of equations $\textbf{E}(\textbf{X})=0$ can be efficiently solved by the Newton-Raphson iteration algorithm, which can be thought as a linear algebra problem of finding the value of $\textbf{X}_{i+1}$ via its value in the previous step $\textbf{X}_{i}$
\begin{equation}
	\textbf{X}_{i+1}=\textbf{X}_{i}-\textbf{M}^{-1}(\textbf{X}_{i})\textbf{E}(\textbf{X}_{i}),\label{eq:N_R}
\end{equation}
where $\textbf{M}=\frac{\delta \textbf{E}}{\delta \textbf{X}}$ is the Jacobian matrix.
The procedure is iterated until the difference $\textbf{X}_{N}-\textbf{X}_{N-1}$ is small enough, which is the condition for $\textbf{X}_{N}$ to be considered a static solution.
In addition, the boundary conditions (\ref{eq:BC}) must be maintained throughout the process.

In order to implement the iteration numerically, we make a coordinate compactification $z=r^{-1}$ such that the radial direction is bounded in $z\in [0,1]$.
Note that the radial position of the horizon is fixed at $r_{h}=1$ using the reparameterization freedom (\ref{eq:shift}).
Discretizing the $z$-coordinate with Chebyshev-Gauss-Lobatto grid points and replacing the radial derivative with corresponding differentiation matrix, the equation (\ref{eq:N_R}) is converted into a series of algebraic operations, which is conveniently implemented with a code library such as numpy.

\subsection{phase diagram}\label{sec:PD}
Due to the confining AdS boundary that restricts the escape of matter, the electric charge and energy of the system are conserved during evolution, indicating that such dynamics essentially occurs in the microcanonical ensemble.
In this case, the relevant thermodynamic potential describing the competitive relationship between the several thermal phases in equilibrium is the entropy, and the dominant thermal phase is the one with the largest entropy.
Therefore, to a certain extent, the microcanonical phase diagram is a key factor in judging the stability of thermal phases in the asymptotically AdS spacetime.

%%%%%%%%%%%%%%%%%%
\begin{figure}
	\begin{center}
		\subfigure[]{\includegraphics[width=.49\linewidth]{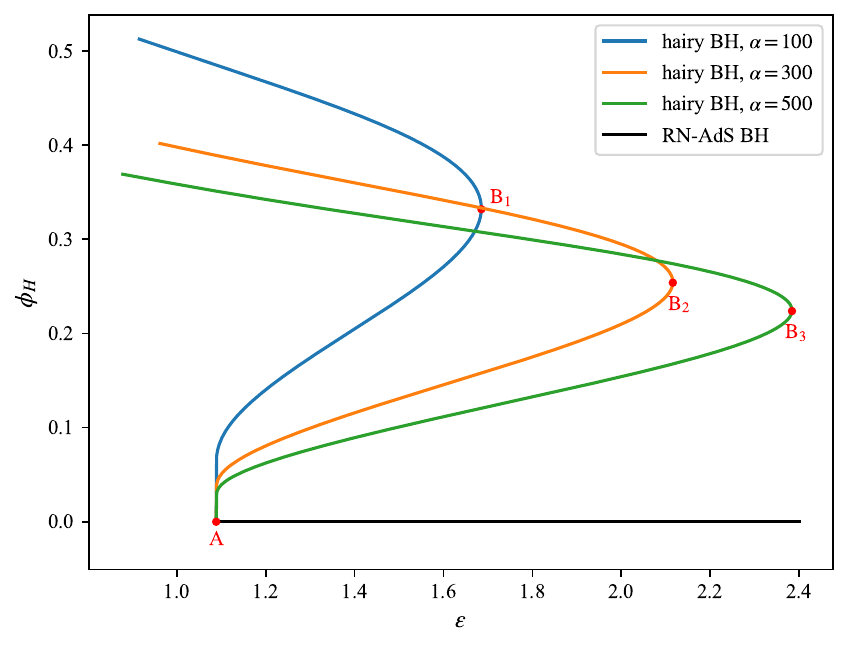}\label{fig:1}}
		\subfigure[]{\includegraphics[width=.49\linewidth]{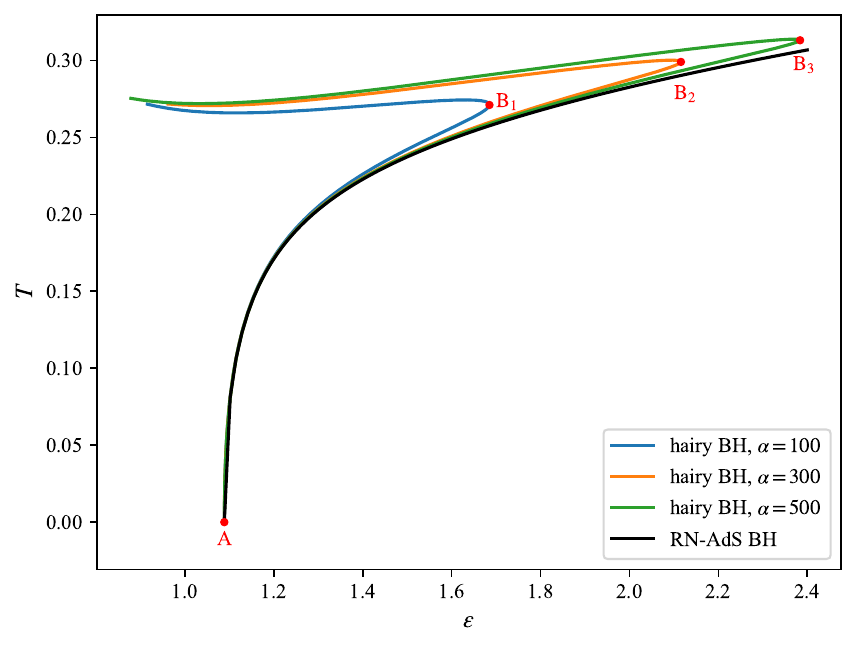}\label{fig:2}}
		\subfigure[]{\includegraphics[width=.49\linewidth]{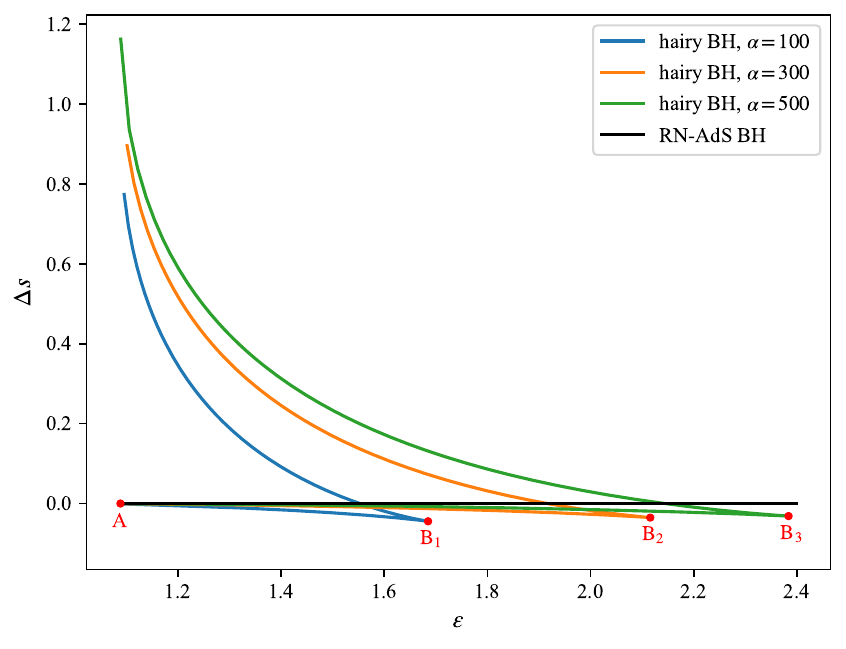}\label{fig:3}}
		\caption{The value of the scalar field at the horizon (a), the temperature (b) and the difference between the entropy density of hairy black holes and RN-AdS black holes (c) as a function of energy density.
		The black curve represents the branch of RN-AdS black holes.
		Other curves of different colors represent branches of hairy black holes with different values of the coupling constant $\alpha$.}
		\label{fig:1-3}
	\end{center}
\end{figure}
%%%%%%%%%%%%%%%%%%

In this model, the microcanonical phase diagrams with different values of coupling constant $\alpha$ are shown in figure \ref{fig:1-3} as the energy density dependence of physical quantities.
As we can see from figure \ref{fig:1}, in addition to RN-AdS black hole solutions with trivial scalar field, for a sufficiently large value of coupling constant $\alpha$, there is also a branch of black hole solutions whose horizon surface is attached to scalar condensation, which connects to the branch of RN-AdS black holes at point $A$ representing the extremal RN-AdS black hole.
Fixing a general value of coupling constant $\alpha$, one can observe that the branch of hairy black holes possesses a turning point $B$, which represents the maximum value of energy density of hairy black hole solutions and depends on the coupling constant $\alpha$.
Such turning point divides the branch of hairy black holes into two parts.
On the one hand, the $AB$ region directly connected to the extremal RN-AdS black hole with zero temperature, usually referred to as the branch of cold hairy black holes, has not only less scalar condensation, but also a lower temperature than the other part as shown in figure \ref{fig:2}.
On the other hand, the region extending from the turning point $B$ to the over-extremal region with decreasing energy is called the branch of hot hairy black holes due to the higher temperature.
With the increase of the coupling constant $\alpha$, both branches of hairy black holes exhibit the same decreasing behavior in terms of scalar condensation.
However, for the temperature, the branch of hot hairy black holes increases while the branch of cold hairy black holes decreases as the coupling constant $\alpha$ increases.

In figure \ref{fig:3}, we would like to reveal the competitive relationship between above three thermal phases in the microcanonical ensemble.
To make the diagram clearer, on the vertical axis we actually plot the difference between the entropy density of hairy black holes and RN-AdS black holes.
As we can see from it, the three branches of solutions form the shape of a swallowtail, among which, the entropy density of the branch of cold hairy black holes is always the smallest.
Due to the second law of black hole mechanics, which requires that the entropy never decreases during the dynamical process, the cold hairy black hole, as an excited state of the system, is expected to be dynamically unstable and spontaneously evolve to a ground state with greater entropy under perturbations.
For a system with fixed energy, both the RN-AdS black hole state and hot hairy black hole state meet the criteria of being the ground state.
In a low-energy region, the state of hot hairy black hole has the largest entropy density and thus acts as the dominant phase.
However, the entropy density of the RN-AdS black hole state gradually exceeds that of the hot hairy black hole state as the energy density of the system exceeds a critical value, becoming the global ground state.
In fact, both of these local ground states (RN-AdS black hole and hot hairy black hole) can serve as the final fate of a dynamically unstable excited state (cold hairy black hole), independent of which is the dominant state with maximum entropy.
This will be verified from the perspective of nonlinear dynamics in section \ref{sec:BD}.
In addition, for an ensemble with a fixed energy, the entropy gap between the hot and cold hairy black holes increases significantly with the increase of the coupling constant $\alpha$, indicating that the dynamical barrier in the excitation process of the hot hairy black hole increases with the parameter $\alpha$.
Interestingly, the entropy gap between the RN-AdS black hole and the cold hairy black hole decreases as the coupling constant $\alpha$ increases, indicating that the RN-AdS black hole is more likely to be excited under the strong coupling condition.
These conclusions are consistent with the numerical simulation results presented in section \ref{sec:CD}.

\section{Stability}\label{sec:S}
In this section, we further investigate the linear stability of the thermal phases obtained in the previous section by the linear perturbation theory.
The corresponding quasinormal mode spectrums are numerically calculated, which show that the hot hairy black hole as the local ground state is linearly stable while the cold hairy black hole as the excited state is dynamically unstable.

%using the generalized eigenvalue method \cite{Jansen:2017oag}, which has been proven to be simple and efficient.

For RN-AdS black holes, the superradiance and near-horizon instabilities are two important mechanisms leading to the dynamical transition to hairy black holes.
For the case of a real scalar field in our work, which cannot extract electric charge from the black hole, the superradiant instability is suppressed.
The near-horizon instability is quite universal and occurs for black holes with extremal configuration, which can be triggered by both charged and neutral scalar fields.
For our model here, such a instability is only possible for a near-extremal RN-AdS black hole in the large black hole limit $r_{h}\rightarrow\infty$, which is not within our consideration.
In what follows, we focus on the linear stability of hot and cold hairy black holes.

\subsection{Effective potential}
From quantum mechanics \cite{buell1995potentials,Nandi:1995gxo}, the effective potential is an important mechanism to characterize the stability of the system.
The emergence of instability requires that the effective potential be negative in some regions.
In this subsection, we reveal the effective potential of the black hole solutions in this model along the radial direction to preliminarily judge their stability.
To this end, we take the following metric ansatz for convenience
\begin{equation}
	ds^{2}=-\widetilde{N}(t,r)e^{-2\widetilde{\Delta}(t,r)}dt^{2}+\frac{1}{\widetilde{N}(t,r)}dr^{2}+r^{2}d\Omega^{2}_{2}.
\end{equation}

Considering a time-dependent linear perturbation with spherical symmetry, the corresponding metric and scalar fields are assumed to be of the form
\begin{subequations}
	\begin{align}
		\widetilde{N}(t,r)&=N(r)+\varepsilon \delta N(r)e^{-i\omega t},\\
		\widetilde{\Delta}(t,r)&=\Delta(r)+\varepsilon \delta \Delta(r)e^{-i\omega t},\\
		\widetilde{\phi}(t,r)&=\phi(r)+\varepsilon \delta \phi(r)e^{-i\omega t}.
	\end{align}\label{eq:4.2}
\end{subequations}
The Maxwell field is determined by equation (\ref{eq:S_E}) as $\partial_{r}\widetilde{A}(t,r)  =\frac{Qe^{-\widetilde{\Delta}}}{r^{2}f(\widetilde{\phi}) }$.
Here the symbol $\varepsilon$ stands for the control parameter of the infinitesimal expansion, and the complex frequency $\omega=\omega_{R}+i\omega_{I}$, also known as quasinormal mode, corresponds to the eigenvalue of the perturbative eigenstate $\{ \delta N,\delta \Delta,\delta \phi\}$.
The configuration of the leading terms $\{ N,\Delta,\phi\}$ is the static background solution of the field equations.
For a mode with a positive imaginary part $\omega_{I}>0$ , one can observe from the ansatz (\ref{eq:4.2}) that the time-dependent solution $\{\widetilde{N},\widetilde{\Delta},\widetilde{\phi}\}$ will exponentially leave the static field configuration in the form $e^{\omega_{I}t}$.
Such a mode is dynamically unstable.
Conversely, a mode with a negative imaginary part $\omega_{I}<0$ decays exponentially, failing to trigger instability in the system.

Substituting the ansatz (\ref{eq:4.2}) into the equations of motion for the metric (\ref{eq:2.2}) and scalar (\ref{eq:S_E}) fields, the leading order of the expansion equations gives rise to the following static field equations
\begin{subequations}
	\begin{align}
		0&= (N  \phi')'+2r^{-1}N  \phi'+\frac{1}{2}rN \phi'^{3}    + 2 \phi +   \frac{Q^{2}}{4r^{4}f^{2}}   \frac{{df}}{d\phi}, \\
		0&= N' + \frac{1}{2}r N \phi'^{2}+ r^{-1}N - r \phi^{2} - 3r   - r^{-1}+\frac{Q^{2}}{4 r^{3} f},\\
		0&=\Delta'+\frac{1}{2}r \phi'^{2},
	\end{align}
\end{subequations}
where prime stands for the derivative with respect to the radius $r$.
Obviously, hot hairy black holes, cold hairy black holes and bald black holes solve the above equations. 
At the subleading order, one can find that the perturbations of the metric fields can be expressed by the perturbation of the scalar field, as follows
\begin{subequations}
	\begin{align}
		0&=\delta N+ r{N} {\phi}'\delta \phi , \\
		0&= \delta \Delta'+r {\phi}'\delta \phi' ,
	\end{align}
\end{subequations}
indicating that the linear order Klein-Gordon equation is decoupled from the perturbation of the metric fields.
Introducing the tortoise coordinate $dr_{*}=e^{\Delta}N^{-1}dr$ and defining a new radial function $\delta\phi=r^{-1}\Psi$, one can extract a Schrodinger-like equation for the perturbation of the scalar field from the subleading order of the Klein-Gordon equation
\begin{equation}
	0=\frac{d^{2}}{dr_{*}^{2}}\Psi+ w^{2} \Psi -V_{\text{eff}}\Psi,\label{eq:4.5}
\end{equation}
with the effective potential
\begin{widetext}
\begin{equation}
	\begin{aligned}
		V_{\text{eff}}=&\frac{Ne^{-2\Delta}}{r^{2}}\left[\left(1+ 3r^{2}\right)\left(1 - r^{2} \phi'^{2}\right) - N   + m^{2}r^{2}\left( \frac{ 1}{2}r^{2} \phi^{2} \phi'^{2} + 2  r \phi \phi' - \frac{ 1}{2}\phi^{2} + 1\right)  \right.\\
		&\left.-\frac{Q^{2}}{4r^{2}f}\left(1 - r^{2} \phi'^{2} +\frac{ {d^{2}f}}{  fd\phi^{2}}+ 2r\phi'\frac{ {df} }{ fd\phi} -2\left(\frac{ {df}}{ fd\phi}\right)^{2}\right)\right].
	\end{aligned}
\end{equation}
\end{widetext}
One can observe that such effective potential vanishes at both the event horizon and the AdS boundary in the case of $m^{2}=-2$, whereas it diverges at the AdS boundary for the case of a massless scalar field.

%%%%%%%%%%%%%%%%%%
\begin{figure}
	\begin{center}
		\subfigure[]{\includegraphics[width=.49\linewidth]{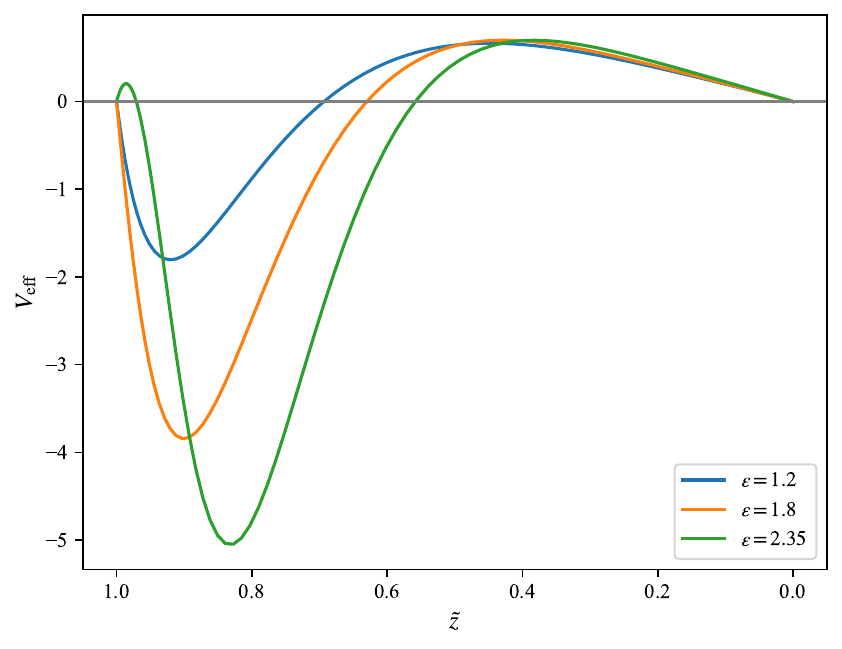}\label{fig:4}}
		\subfigure[]{\includegraphics[width=.49\linewidth]{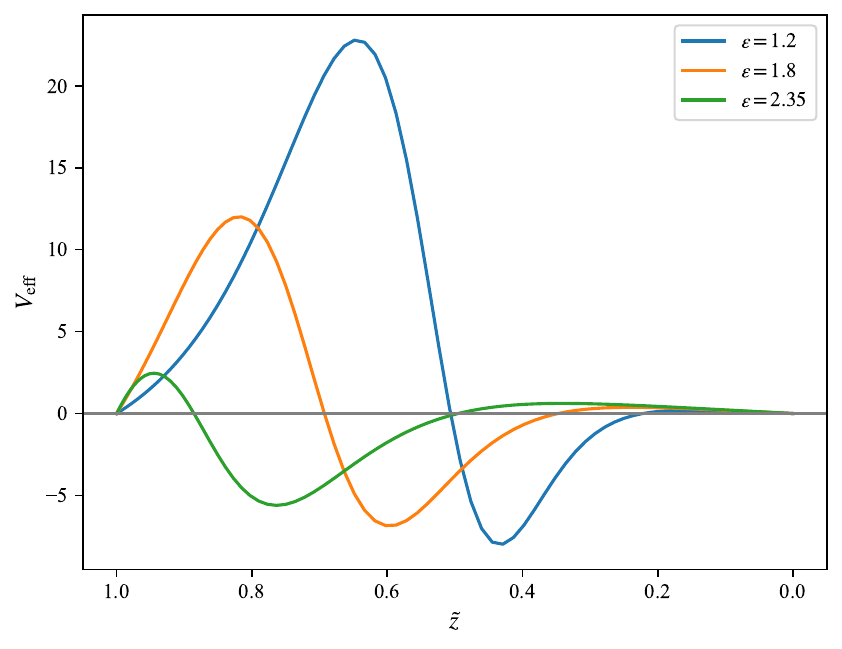}\label{fig:5}}
		\subfigure[]{\includegraphics[width=.49\linewidth]{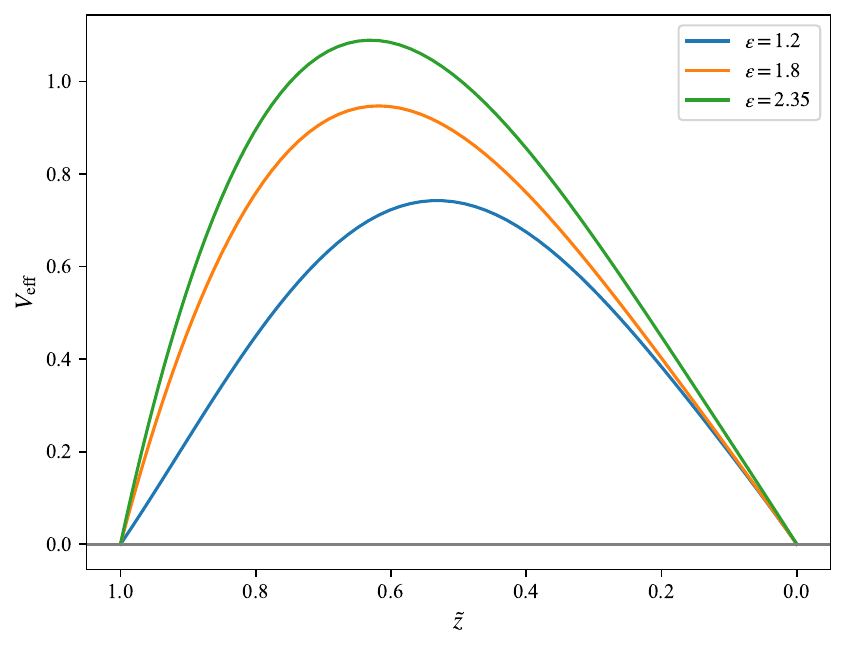}\label{fig:6}}
		\subfigure[]{\includegraphics[width=.49\linewidth]{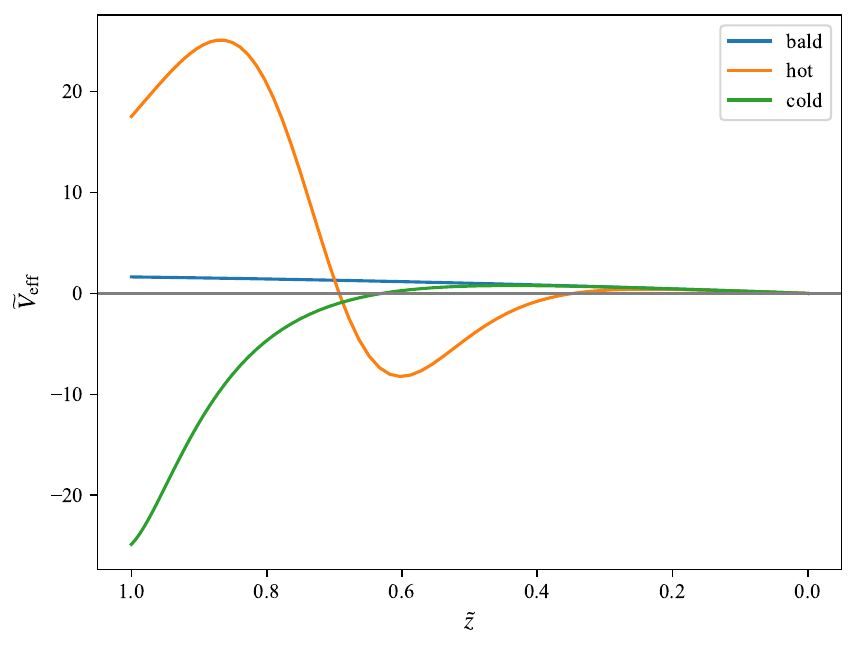}\label{fig:7}}
		\caption{The effective potentials of cold hairy black holes (a), hot hairy black holes (b) and RN-AdS black holes (c) with different energy densities as a function of the rescaled radial compactified coordinate $\tilde{z}=r_{h}/r$, where $r_{h}$ represents the radius of the event horizon.
		The coupling constant is fixed to $\alpha=500$.
		(d): The rescaled effective potentials of the cold hairy black hole, hot hairy black hole and RN-AdS black hole with energy density $\epsilon=1.8$.}
		\label{fig:4-7}
	\end{center}
\end{figure}
%%%%%%%%%%%%%%%%%%

The resulting effective potentials for the cold hairy black holes, hot hairy black holes and RN-AdS black holes are shown in figures \ref{fig:4}, \ref{fig:5} and \ref{fig:6}, respectively.
Since the RN-AdS black hole solutions have a positive definite effective potential, they are free from radial instability.
However, for the hairy black hole solutions, it turns out that the effective potential always has a negative region.
The difference is where the negative region exists.
On the one hand, for solutions in the cold branch, such negative region comes into play near the event horizon and gradually intensifies along the direction of the cold branch from the connection point with the branch of RN-AdS black holes (point $A$ representing the extremal RN-AdS black hole) to the bifurcation point of the branch of hairy black holes (point $B$), that is, the direction in which the energy density increases.
As the bifurcation point is approached, a positive region develops near the event horizon.
On the other hand, as the configuration smoothly transitions from the cold branch to the hot branch through the bifurcation point, the positive region near the event horizon gradually grows.
Along the direction of the hot branch away from the bifurcation point, such positive region is significantly enlarged and gradually plays a dominant role as the energy density of the system decreases.
The negative region can only move away from the event horizon towards the AdS boundary.
As a result, near the event horizon, a hot hairy black hole exhibits a potential barrier while a cold hairy black hole possesses a potential well.

According to the results of quantum mechanics \cite{buell1995potentials}, for a one-dimensional potential, the existence of bound states capable of triggering instability requires that the integral of the effective potential over the entire space is negative.
In order to compare with it, we introduce the rescaled effective potential defined as
\begin{equation}
	\widetilde{V}_{\text{eff}}=r_{h}z^{-2}e^{\Delta}N^{-1}V_{\text{eff}},
\end{equation}
such that $\int_{-\infty}^{+\infty}V_{\text{eff}}dr_{*}=\int_{0}^{1}\widetilde{V}_{\text{eff}}dz$.
Without loss of generality, taking the ensemble with energy density of $\epsilon=1.8$ as an example, the profiles of the corresponding rescaled effective potential of the three types of black holes are shown in figure \ref{fig:7}.
It turns out that only the integral of the effective potential of a cold hairy black hole is negative, and thus is expected to be dynamically unstable.
We have verified that such integral is negative for the entire branch of cold hairy black holes.
However, these qualitative analyses still cannot give definitive evidence of instability.
To this end, one needs to obtain the eigenvalue $\omega$ of the linear perturbation to determine whether there is an unstable mode with a positive imaginary part.

\subsection{Quasi-normal modes}\label{sec:QNM}
In this subsection, we numerically solve the quasinormal spectrum to give direct evidence of instability quantitatively.
Since the configurations of hairy black holes are obtained by numerical method, the generalized eigenvalue method \cite{Jansen:2017oag} is used, which is simple and efficient for this case.
By discretizing the field configurations of the static background solution with a pseudospectrum, this method converts the solving process of the equation (\ref{eq:4.5}) into a generalized eigenvalue problem.
The complex frequency $\omega$ to be calculated is the corresponding eigenvalue.

%%%%%%%%%%%%%%%%%%
\begin{figure}
	\begin{center}
		\subfigure[]{\includegraphics[width=.49\linewidth]{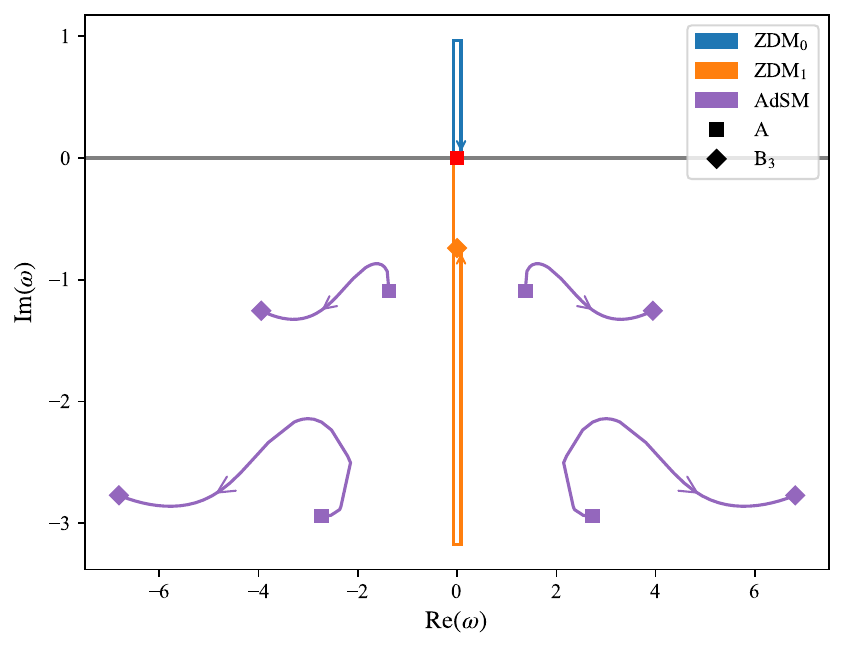}\label{fig:8}}
		\subfigure[]{\includegraphics[width=.49\linewidth]{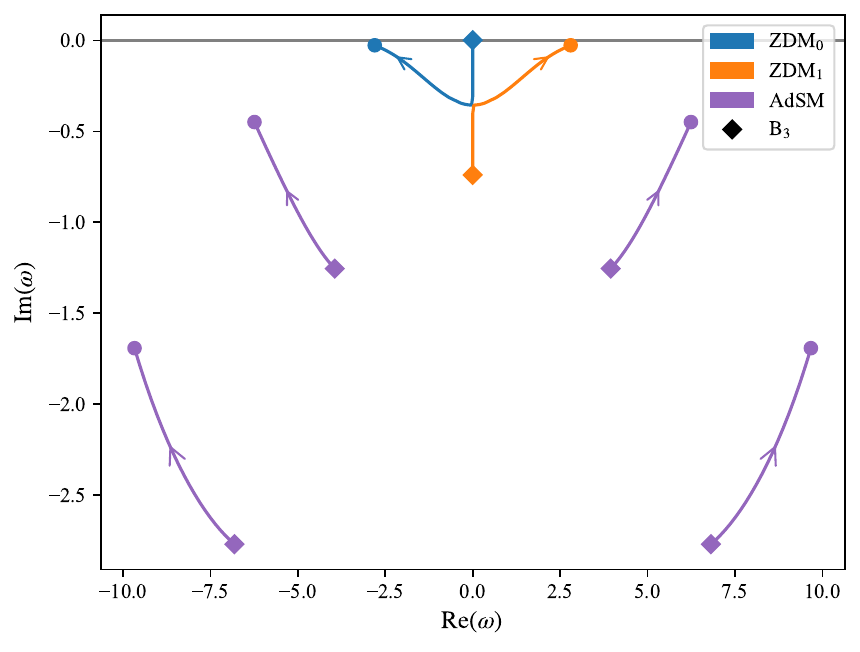}\label{fig:9}}
		\subfigure[]{\includegraphics[width=.49\linewidth]{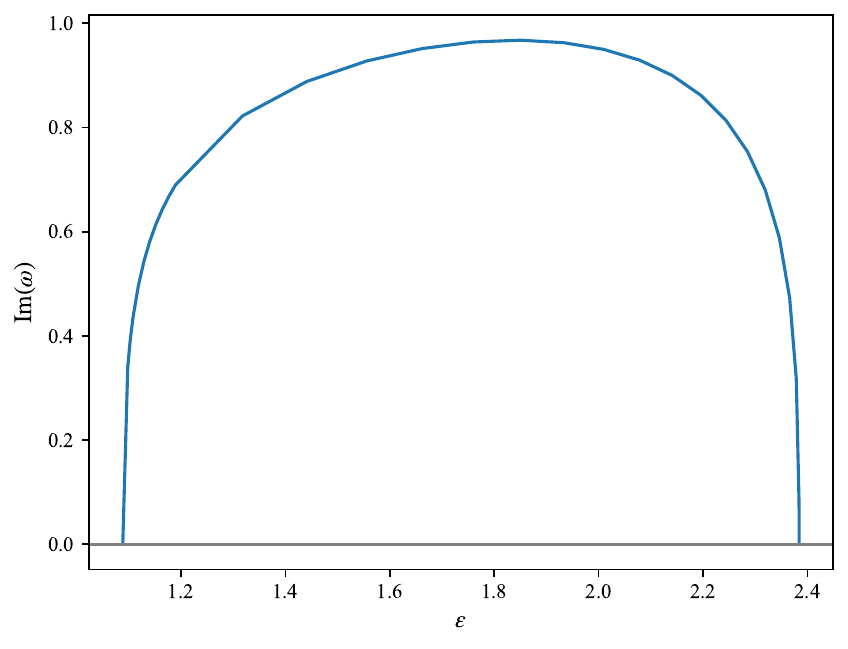}\label{fig:10}}
		\caption{The trajectories of the first four AdS modes (purple) and the first (blue) and second (orange) zero-damped modes along the branches of cold (a) and hot (b) hairy black holes.
				(a): Arrows indicate the direction from the connection point (point $A$) between the branches of hairy and RN-AdS black holes to the bifurcation point (point $B$) of the branch of hairy black holes.
				(b): Arrows indicate the direction of decreasing energy from the bifurcation point (point $B$) to the extremal region.
				(c): The imaginary part of the unstable zero-damped mode of the cold branch as a function of the energy density of the system.}\label{fig:8-10}
	\end{center}
\end{figure}
%%%%%%%%%%%%%%%%%%

The resulting quasinormal spectrums of the cold and hot hairy black holes are shown in figures \ref{fig:8} and \ref{fig:9}, respectively.
The purple dots represent the AdS modes, which dominates in the small black hole limit $r_{h}\rightarrow 0$.
The blue and orange dots represent the zero-damped modes, which converge at the origin of the complex plane as a branch point in the case of an extremal RN-AdS black hole \cite{Yang:2012pj,Zimmerman:2015trm,Zimmerman:2016qtn}.
For the branch of RN-AdS black holes, all the modes are located on or below the real axis, indicating the dynamical linear stability.
However, along the branch of cold hairy black holes from the connection point $A$ representing the extremal RN-AdS black hole to the bifurcation point $B$, one of the zero-damped modes gradually climbs upward along the imaginary axis from the origin, while the others move down to the lower half of the imaginary axis.
Such a mode with a positive imaginary part represents the occurrence of dynamical linear instability.
After the imaginary part of the single unstable mode reaches a maximum value, the linear instability of the system gradually weakens as it decreases.
Until the bifurcation point $B$ is reached, the imaginary part of this single unstable mode returns to the origin, indicating that the instability completely disappears.
We reveal more clearly in figure \ref{fig:10} that the positive imaginary part of the single unstable mode varies with the energy density in this process.
Along the branch of hot hairy black holes from the bifurcation point $B$ to the over-extremal region, on the other hand, the dominant mode initially moves downwards from the origin along the imaginary axis, and then gradually approaches the real axis again after reaching a turning point, accompanied by the growth of the real part. 
With the approach of the extremal configuration with zero horizon area, this mode gradually converges to the real axis.
Therefore, from the above results, we can conclude that only the cold hairy black holes are dynamically linearly unstable, and there is only a single unstable mode without real part.

\section{Dynamics of the local ground state}\label{sec:CD}
\begin{figure}
	\begin{center}
		\includegraphics[width=.6\linewidth]{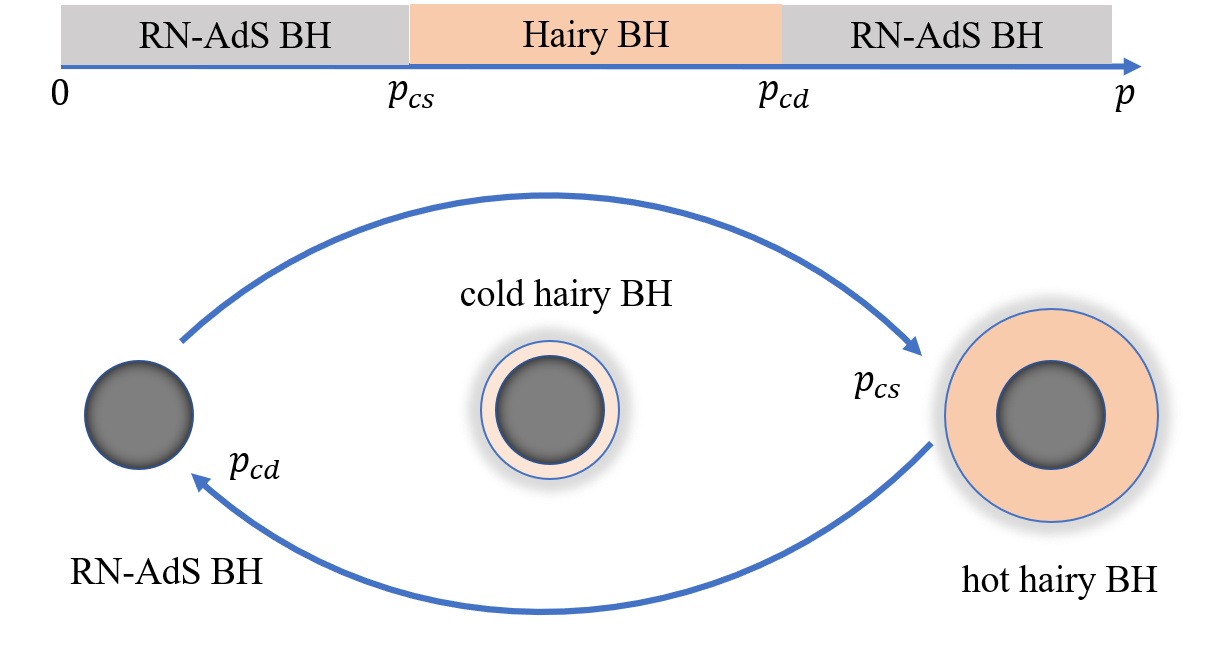}
		\caption{The schematic diagram of the continuous accretion process of the scalar field to a central black hole (BH). The accretion strength is denoted by the symbol $p$. The values $p_{cs}$ and $p_{cd}$ correspond to the critical scalarization and critical descalarization processes, respectively.}
		\label{fig:schematic}
	\end{center}
\end{figure}

From the results of the linear perturbation theory in the previous section, the RN-AdS black holes and hot hairy black holes are dynamically linearly stable, acting as two local ground states.
In this section, by numerically solving the time-dependent field equations, we simulate the fully nonlinear accretion process of a scalar field to a central black hole to reveal the real-time dynamics during the excitation process of the ground state, where scalarization or descalarization phenomenon occurs depending on the central black hole.
The corresponding schematic diagram of such continuous accretion process is shown in figure \ref{fig:schematic}.
Due to the linear stability of the RN-AdS and hot hairy black holes, the disturbance of the accretion process of small strength, which can only increase the energy of the system without changing its essential properties, gradually dissipates in the background spacetime.
Until the accretion strength exceeds a threshold, the RN-AdS and hot hairy black holes with nonlinear instability are dynamically interconverted by crossing a cold hairy black hole with linear instability.
From the microcanonical phase diagram in figure \ref{fig:1-3}, it can be seen that the branch of hot hairy black holes ends at point $B$, indicating that the accretion process with sufficient strength, which brings in enough energy, can always make the system converge to the state of RN-AdS black hole.
The real-time dynamics of the above physical processes will be revealed in detail in the following.

%%%%%%%%%%%%%%%%%%
%%%%%%%%%%%%%%%%%%

Among them, it is particularly important to emphasize the critical dynamics at the strength threshold, near which the evolved system tends to converge to a certain critical state in the dynamical intermediate process.
In particular, if such critical state has only one unstable mode, then we can obtain it by parameterizing the initial value with a single parameter and fine-tuning the characteristic parameter to reach a critical value.
At the late time of critical evolution, the system will enter the linear region of the critical solution, which can be effectively approximated by 
\begin{equation}
	\phi(t,r)\approx\phi_{*}(r)+(p-p_{*})e^{-i\omega _{*}t}\delta\phi(r)+\text{decaying modes}.\label{eq:linear_critical}
\end{equation}
Here $\phi_{*}(r)$ represents the static configuration of the critical state, and $\delta\phi(r)$ is the only unstable eigenmode associated with the eigenvalue $\omega _{*}$, which has a positive imaginary part.
On the one hand, for the case where the parameter $p$ is exactly equal to the critical value $p_{*}$, the only unstable mode cannot be triggered, causing the system to permanently stay in the critical state.
However, on the other hand, for the parameter $p$ slightly away from the critical value $p_{*}$, such an unstable mode will grow exponentially in the later stage of evolution and push the system away from the critical state to reach the final stable state.
Interestingly, when the parameter $p$ leaves the critical value $p_{*}$ in different directions, the final state often has distinct essential properties.
Furthermore, the time that the system stays on the critical state during the dynamical intermediate process satisfies
\begin{equation}
	\tau\propto -\text{Im}[\omega _{*}]^{-1}\text{ln}(|p-p_{*}|),\label{eq:5.2}
\end{equation}
where $\text{Im}[\omega _{*}]$ stands for the imaginary part of the eigenvalue $\omega _{*}$.
Such a relationship is obtained by requiring that the coefficient of the unstable mode in equation (\ref{eq:linear_critical}) grows to a finite size, $|(p-p_{*})e^{-i\omega _{*}\tau}|\sim O(1)$, which represents the end point of the linear region of the intermediate critical solution.

\subsection{Critical scalarization}
In this subsection, we focus on the dynamical accretion process of a scalar field towards a central RN-AdS black hole with coupling constant $\alpha=500$.
That is, the physical process near the critical point $p_{cs}$ in figure \ref{fig:schematic}.
Without loss of generality, we choose a seed RN-AdS black hole with energy density $\epsilon=1.6$ as the initial configuration and impose a scalar field perturbation of the form
\begin{equation}
	\delta\phi=pz^{2}(1-z)^{2}e^{-w(z-z_{c})^{2}},\label{eq:5.3}
\end{equation}
with the radial compactified coordinate $z=r^{-1}$.
The width and center position of the Gaussian function are fixed as $w=50$ and $z_{c}=0.5$.
Since the apparent horizon is fixed at the radial position $z_{h}=1$ by using the reparameterization freedom (\ref{eq:shift}), such form of perturbation characterizes a Gaussian-type scalar field distributed outside the central black hole.
The energy of the system increases with the increase of the perturbation amplitude $p$, so this process can be regarded as the accretion process of the scalar field to a central black hole, with the accretion strength $p$.

%%%%%%%%%%%%%%%%%%
\begin{figure}
	\begin{center}
		\subfigure[]{\includegraphics[width=.49\linewidth]{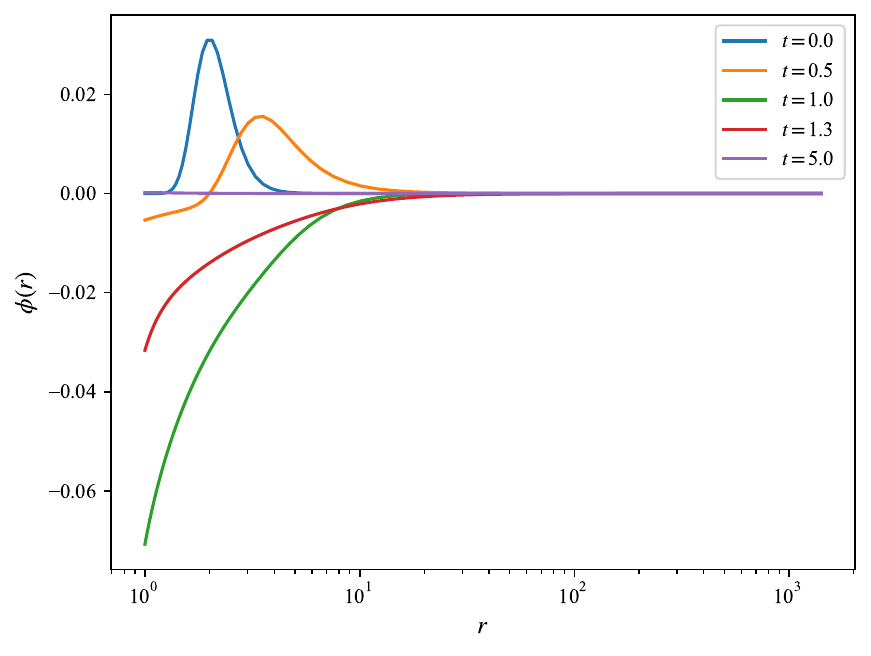}\label{fig:12}}
		\subfigure[]{\includegraphics[width=.49\linewidth]{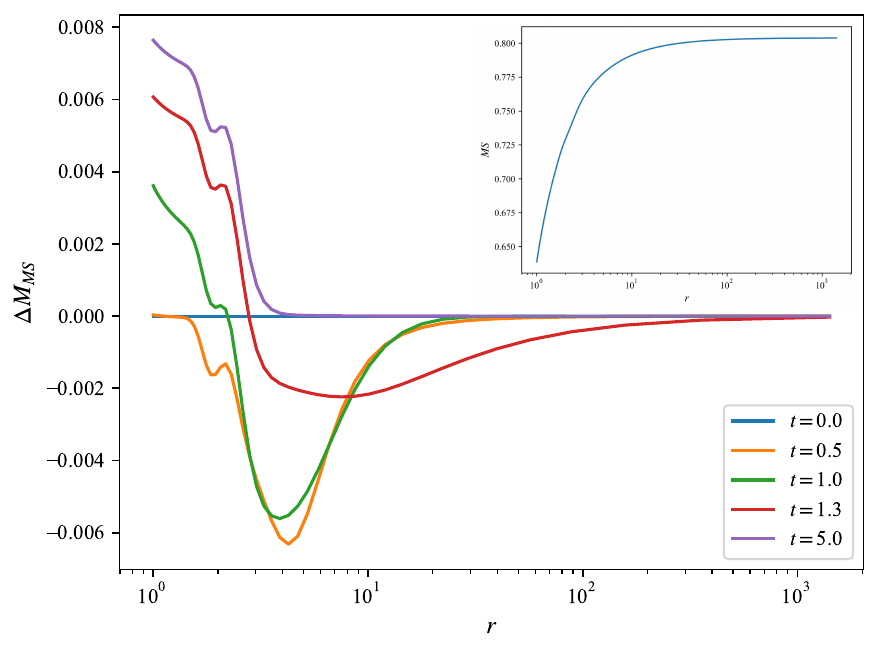}\label{fig:13}}
		\subfigure[]{\includegraphics[width=.49\linewidth]{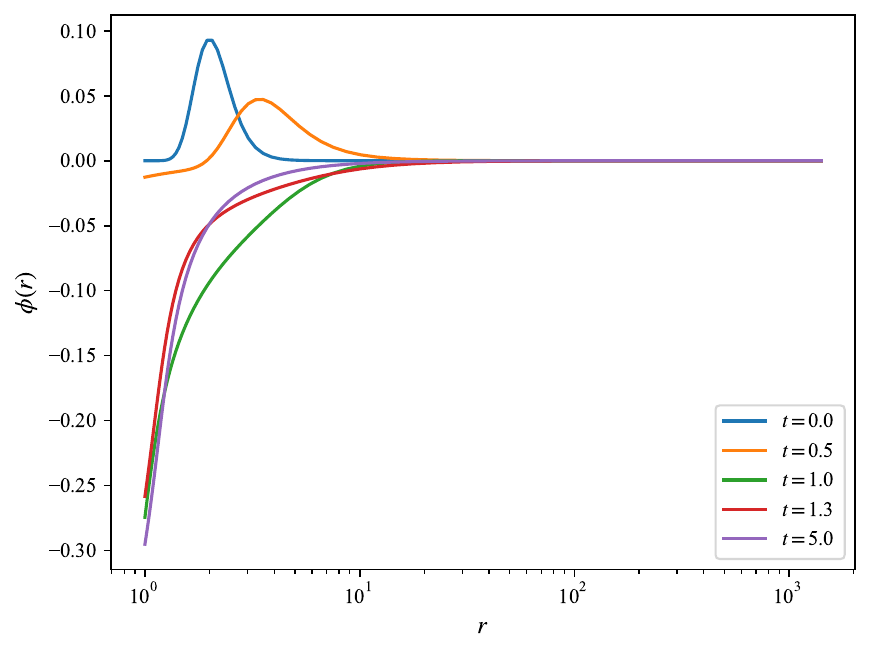}\label{fig:14}}
		\subfigure[]{\includegraphics[width=.49\linewidth]{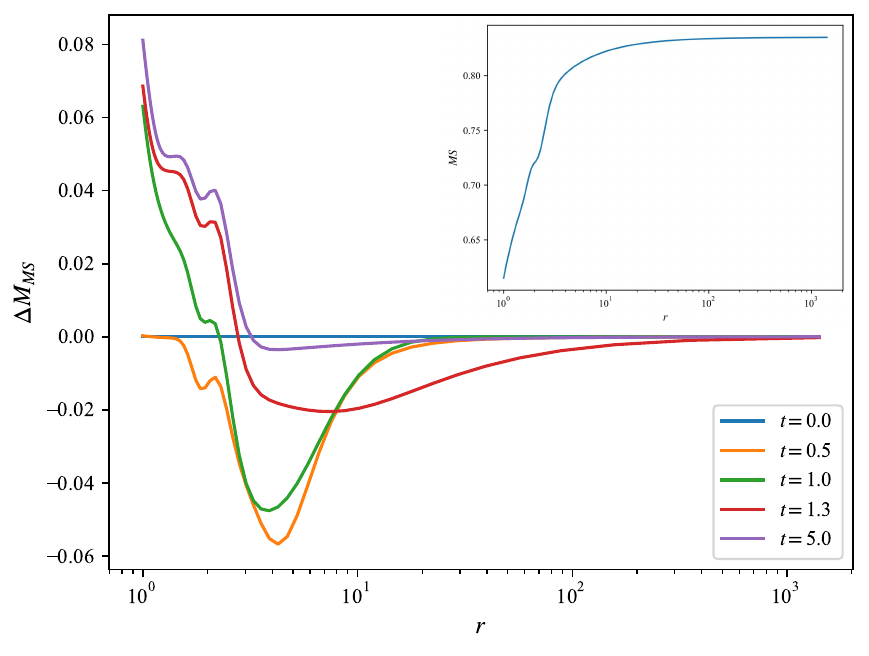}\label{fig:15}}
		\caption{The profile of the scalar field (a, c) and the MS mass (b, d) at different times. 
				The upper and lower panels correspond to the cases of small disturbance $p=0.5$ and large disturbance $p=1.5$, respectively.
				In order to describe the variation of the quasi-local mass more clearly, we actually show the difference of the MS mass at different moments from the initial moment, where the MS mass distribution of the initial data is shown in the inset.}\label{fig:12-15}
	\end{center}
\end{figure}
%%%%%%%%%%%%%%%%%%

The evolution of the scalar field configuration and the MS mass during the dynamical accretion process is shown in figure \ref{fig:12-15}.
In the early stages of evolution, the system exhibits similar dynamical behaviors for accretion processes of different strengths.
In the first stage ($t<0.5$), the outgoing scalar field carrying energy propagates towards the AdS boundary, resulting in a decrease in the local mass at the position of the wave packet at the initial moment, as shown in figures \ref{fig:13} and \ref{fig:15}.
Note that the value of the MS mass at the radial coordinate $r$ represents the integral of the energy within radius $r$.
Since the local mass change in the dynamical process is not significant compared to the overall energy, we show the difference between the MS mass at different times and the initial time so that the energy flow is more obvious.
The MS mass at any time converges to a constant on the AdS boundary, which is equal to the ADM mass, indicating that the total mass of the system in the asymptotically AdS spacetime is conserved during the dynamical process.
Then in the second stage ($0.5<t<1$), due to the gravitational potential of the AdS spacetime, the outwardly propagating scalar field is bounced and clustered around the horizon of the central black hole, resulting in a significant increase in the local mass near the horizon.
Interestingly, the subsequent fate of the scalar field depends on the specific accretion strength.
On the one hand, for a weak accretion strength, such as the accretion process of $p=0.5$ shown in figures \ref{fig:12} and \ref{fig:13}, the scalar field is gradually absorbed by the central black hole in the later stages of evolution ($1<t$), leaving a bald black hole with greater energy.
Since the energy carried by the initial scalar field enters the interior of the central black hole, the value of the MS mass is most significantly improved at the horizon compared with the initial value, and then rapidly decreases at the original wave packet of the scalar field.
Obviously, for a radial position outside the original wave packet, the MS mass of the initial and final states remains the same.
On the other hand, for a strong accretion strength $p=1.5$, as shown in figures \ref{fig:14} and \ref{fig:15}, the scalar field eventually converges to a nontrivial configuration, leading to the appearance of a black hole with the scalar condensation attached at the horizon.
Such a final black hole is exactly the hot hairy black hole obtained in the subsection \ref{sec:PD}, and it has been proved to be linearly dynamically stable in the subsection \ref{sec:QNM}.
That is to say, a scalar field accretion process of sufficient intensity can induce a dynamical transition from a linearly stable RN-AdS black hole to a hot hairy black hole.
From the distribution of the MS mass of the final state in figure \ref{fig:15}, it can be seen that most of the energy still enters the interior of the central black hole, resulting in a significant increase in the value of the MS mass at the horizon.
The scalar hair also carries part of the energy, distributed in the bulk.

%%%%%%%%%%%%%%%%%%
\begin{figure}
	\begin{center}
		\subfigure[]{\includegraphics[width=.49\linewidth]{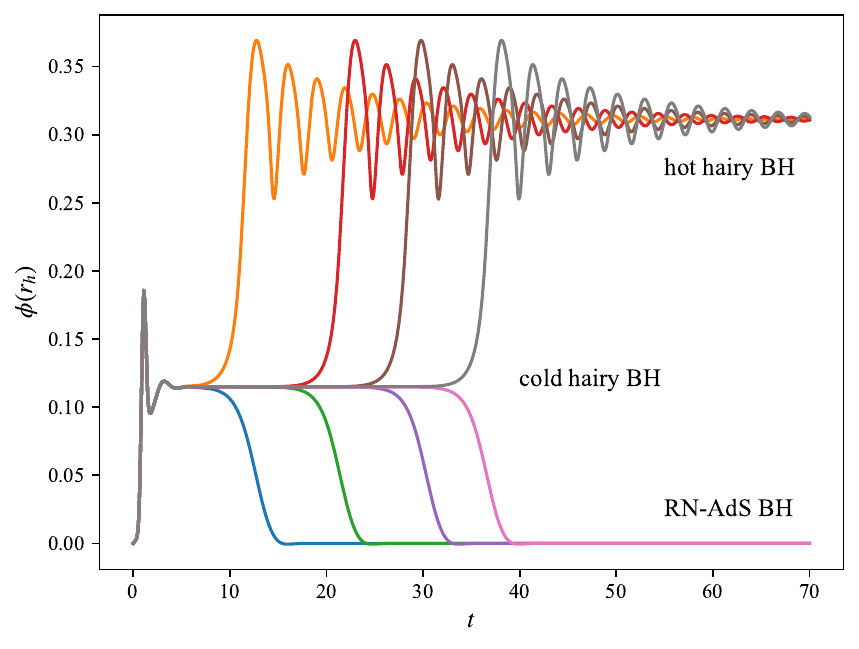}\label{fig:16}}
		\subfigure[]{\includegraphics[width=.49\linewidth]{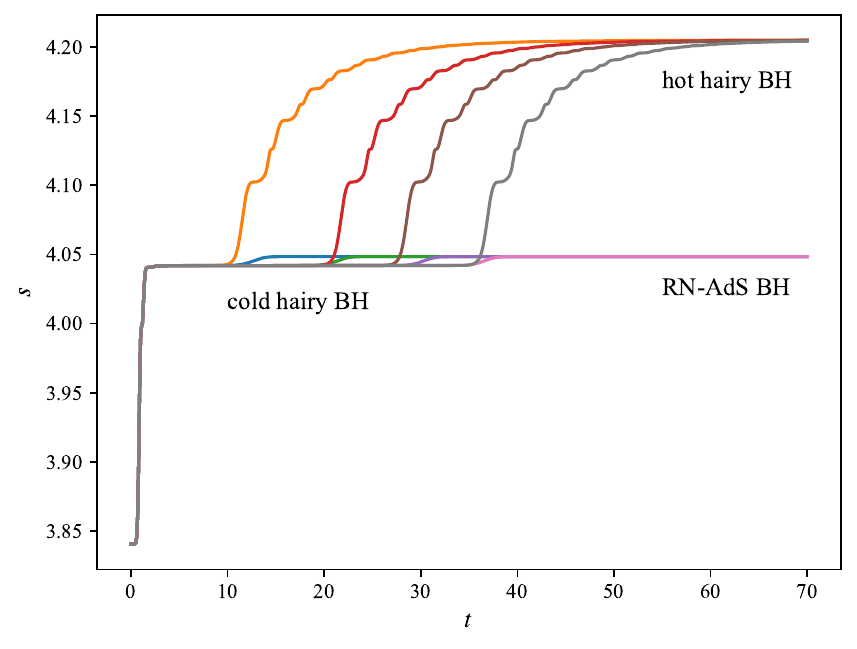}\label{fig:17}}
		\subfigure[]{\includegraphics[width=.49\linewidth]{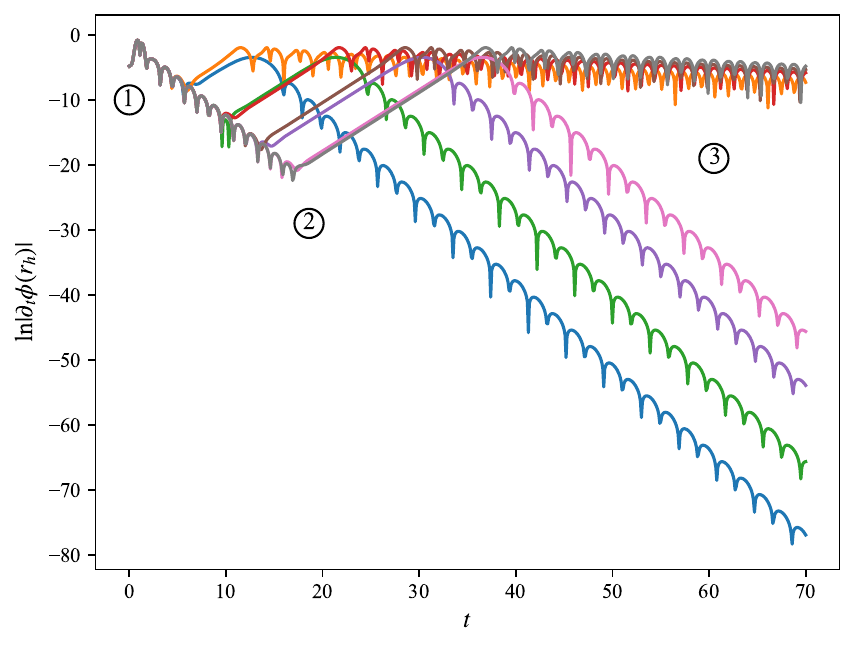}\label{fig:18}}
		\subfigure[]{\includegraphics[width=.49\linewidth]{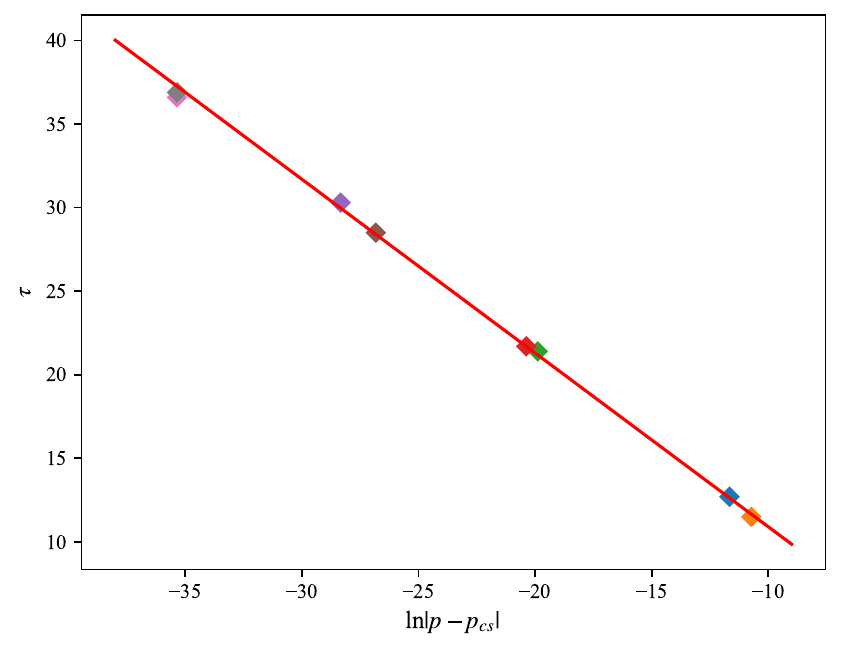}\label{fig:19}}
		\caption{The value of the scalar field at the horizon (a), the entropy density (b) and the value of ln$|\partial_{t}\phi(r_{h})|$ (c) as a function of time in the case where the accretion strength $p$ is close to the critical value $p_{cs}$. 
				(d):  The time $\tau$ of the intermediate solution that stays near the cold hairy black hole with respect to ln$|p-p_{cs}|$.
				All the curves and points of the same color in the figure correspond to each other.}\label{fig:16-19}
	\end{center}
\end{figure}
%%%%%%%%%%%%%%%%%%

The above real-time dynamical processes indicate that there should be a critical value for accretion strength $p$ between $0.5$ and $1.5$ to distinguish two different final states.
A natural question is what kind of dynamical behavior the system exhibits near the critical value of accretion strength $p_{cs}$.
By dichotomy, we keep approaching the critical value and show the numerical results in figure \ref{fig:16-19}.
The evolution of the scalar condensation at the horizon and the entropy density are presented in figures \ref{fig:16} and \ref{fig:17}, respectively, from which it can be seen that a critical state appears in the dynamical intermediate process.
All initial values parameterized by the accretion strength $p$ close to the critical value $p_{cs}$ are attracted to a critical black hole with scalar hair, manifested by the convergence of the scalar field to a static nontrivial configuration.
At the same time, the entropy of the system also stops growing and presents a plateau.
Subsequently, for the dynamical process with the accretion strength greater than the critical value $p_{cs}$, the scalar condensation continues to grow and eventually converges to another static configuration.
The entropy also increases significantly to another constant.
The stable final state of the process is a hot hairy black hole.
The fast oscillating behavior of the scalar condensation in the later stages of evolution is caused by the non-zero real part of the stable dominant mode of the hot hairy black hole.
On the contrary, for the dynamical process with the accretion strength less than the critical value $p_{cs}$, the corresponding scalar condensation decays rapidly after leaving the critical configuration, indicating that the system eventually evolves into an RN-AdS black hole.
These dynamical processes are accompanied by a small increase in entropy, which is guaranteed by the second law of black hole mechanics.
Furthermore, one can observe that the closer the accretion strength is to the critical value $p_{cs}$, the longer the system stays in the critical state during the dynamical intermediate process.
Therefore, it can be inferred that the critical accretion strength $p_{cs}$ just corresponds to the critical black hole.
After verification, such a critical black hole is exactly the linearly unstable cold black hole obtained in the subsection \ref{sec:PD}.

In order to reveal the behavior of the dominant mode in the dynamical process, we show the evolution of the value of ln$|\partial_{t}\phi(r_{h})|$ over time in figure \ref{fig:18}.
One can observe that the whole dynamical process can be divided into three stages.
The short-lived first stage describes the process where the initial values are attracted to a critical black hole, depending on the form of the disturbance.
After that, the evolution system enters the linear region of the critical state, at which point it can be approximated by equation (\ref{eq:linear_critical}).
At the beginning of the second stage, the dynamical process is dominated by the decay modes of the critical black hole.
Due to the deviation between the actual accretion strength and the critical value, the unstable mode will grow exponentially at later times and gradually takes over the evolution process, pushing the system away from the critical state.
The growth exponent can be extracted from the slope in figure \ref{fig:18}, which is equal to the imaginary part of the unstable mode.
In the third stage, the intermediate solution converges to the RN-AdS black hole in the case of subcritical accretion strength and to hot hairy black hole in the case of supercritical accretion strength. 
From the quasinormal modes of hot hairy black holes shown in figure \ref{fig:9}, the dominant stable mode has a small imaginary part, implying a slow decay rate.
Since the critical black hole (cold hairy black hole) emerging in the dynamical intermediate process possesses a single unstable mode shown in figure \ref{fig:8}, the ralationship (\ref{eq:5.2}) is checked in figure \ref{fig:19}, where the slope of the red line is exactly the reciprocal of the imaginary part of the unstable mode.

%%%%%%%%%%%%%%%%%%
\begin{figure}
	\begin{center}
		\includegraphics[width=.5\linewidth]{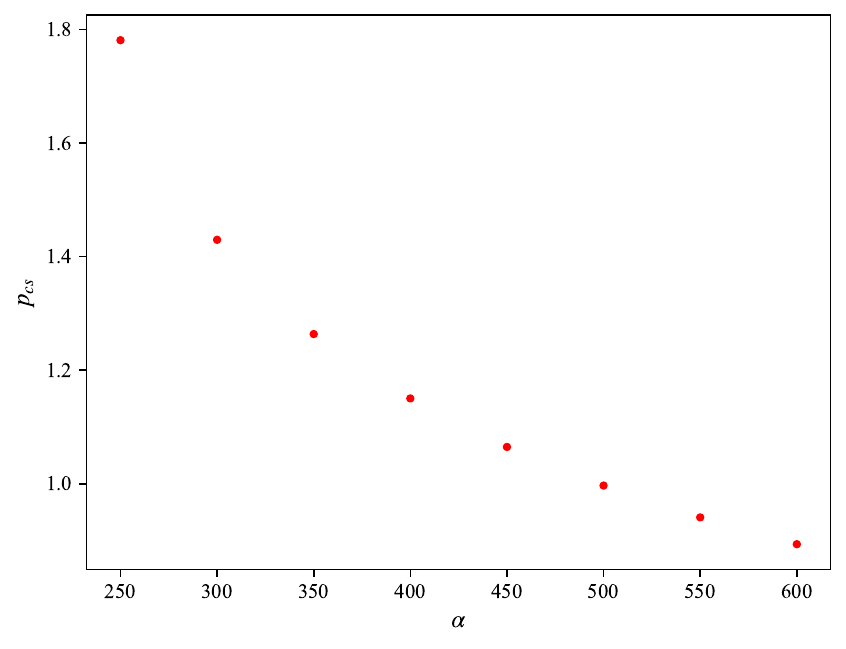}
		\caption{The critical value of accretion strength $p_{cs}$ as a function of the coupling constant $\alpha$.
				The energy density of the central RN-AdS black hole is fixed to be $\epsilon=1.6$.}
		\label{fig:20}
	\end{center}
\end{figure}
%%%%%%%%%%%%%%%%%%

From the microcanonical phase diagram shown in figure \ref{fig:3}, due to the reduced entropy gap between the RN-AdS and cold hairy black holes, one can deduce that the dynamical barrier for the transition from an RN-AdS black hole to a hot hairy black hole gradually decreases with the increase of the coupling constant $\alpha$.
We show in figure \ref{fig:20} the critical accretion strength $p_{cs}$ required to trigger the dynamical transition of the central RN-AdS black hole for different values of the coupling constant $\alpha$, where the monotonically decreasing behavior of the critical accretion strength verifies the inference from the phase diagram.

%%%%%%%%%%%%%%%%%%
\begin{figure}
	\begin{center}
		\subfigure[]{\includegraphics[width=.49\linewidth]{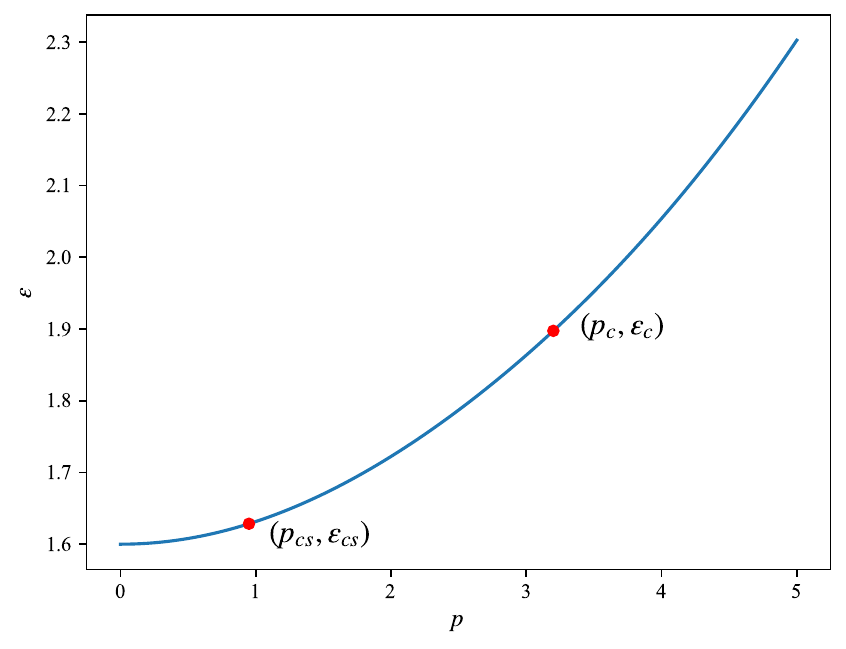}\label{fig:30}}
		\subfigure[]{\includegraphics[width=.49\linewidth]{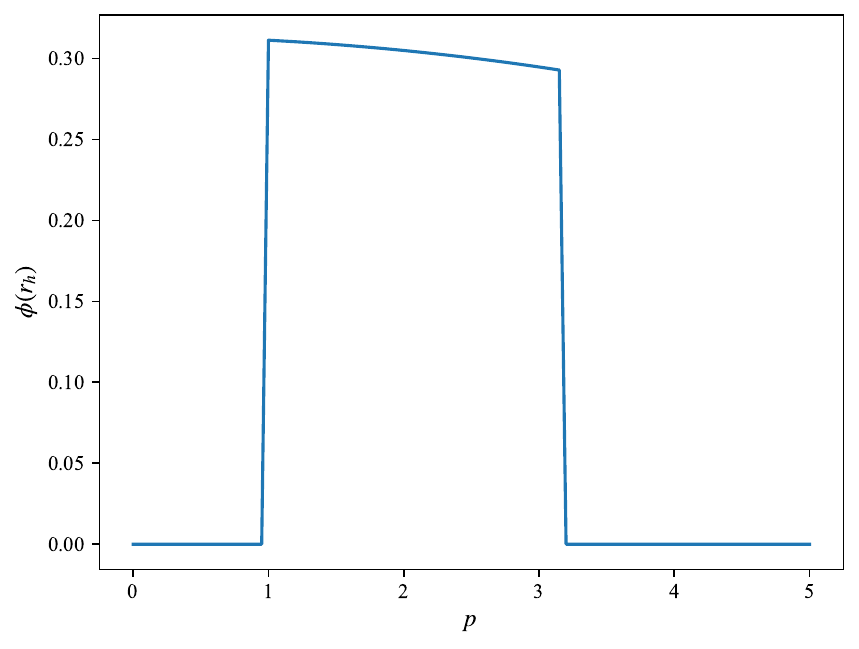}\label{fig:31}}
		\caption{The energy density (a) and the value of the scalar field at the horizon (b) of the final state of the evolution as a function of the accretion strength.}\label{fig:30-31}
	\end{center}
\end{figure}
%%%%%%%%%%%%%%%%%%

The energy density of the gravitational system increases continuously with the further accretion of the scalar field, showing a monotonically increasing behavior with the accretion strength $p$, as shown in figure \ref{fig:30}.
At the first threshold of the accretion strength $p_{cs}$, the value of the final scalar field at the horizon jumps, as shown in figure \ref{fig:31}, indicating a dynamical transition from an RN-AdS black hole to a hot hairy black hole.
Subsequently, such value gradually decreases with the increase of the accretion strength, which is consistent with the result shown in figure \ref{fig:1} that the scalar condensation attached to the event horizon of the hot hairy black hole decreases with the increase of the energy density of the system.
Since the branch of hot hairy black holes ends at the bifurcation point $B$, the system must return to the branch of RN-AdS black holes when the energy density exceeds that of point $B$.
Interestingly, from the dynamical results in figure \ref{fig:31}, instead of passing through the whole branch of hot hairy black holes, the gravitational system evolves to the branch of RN-AdS black holes through a critical dynamical transition in advance, manifested by another jump in the value of $\phi(r_{h})$ at the second threshold of the accretion strength $p_{c}$.
The energy density corresponding to this threshold $p_{c}$ is between those corresponding to the first threshold and the bifurcation point.
The real-time dynamics near the second threshold is similar to the scalar field accretion process towards a hot hairy black hole, which will be revealed in detail in the next subsection.

As a conclusion, the accretion process of a scalar field to a central RN-AdS black hole can trigger its nonlinear instability.
There is a threshold of the accretion strength that induces the dynamical transition from one local ground state (RN-AdS black hole) to the other (hot hairy black hole).
Near the threshold, the system stays on an excited state (cold hairy black hole), which acts as a dynamical barrier for the transition process.
Such a dynamical barrier decreases monotonically with the increase of the coupling constant $\alpha$.
With the continuous accretion of the scalar field, due to the upper limit of the energy density for the domain of existence of the hot hairy black holes, a second threshold of the accretion strength appears, beyond which the gravitational system undergoes a critical dynamical transition and returns to the state of RN-AdS black hole.

\subsection{Critical descalarization}
The dynamical simulation results in the previous subsection show that even if a gravitational system is dynamically stable at the linear level, it can still transition to another linearly stable local ground state under sufficiently large disturbance.
In this subsection, by simulating the dynamical accretion process of a scalar field to a central hot hairy black hole, we reveal that the transition process between local ground states is bidirectional.
Without loss of generality, we take the hot hairy black hole under the same ensemble with energy density $\epsilon=1.6$ as the initial value and impose the same form of disturbance (\ref{eq:5.3}).

%%%%%%%%%%%%%%%%%%
\begin{figure}
	\begin{center}
		\subfigure[]{\includegraphics[width=.49\linewidth]{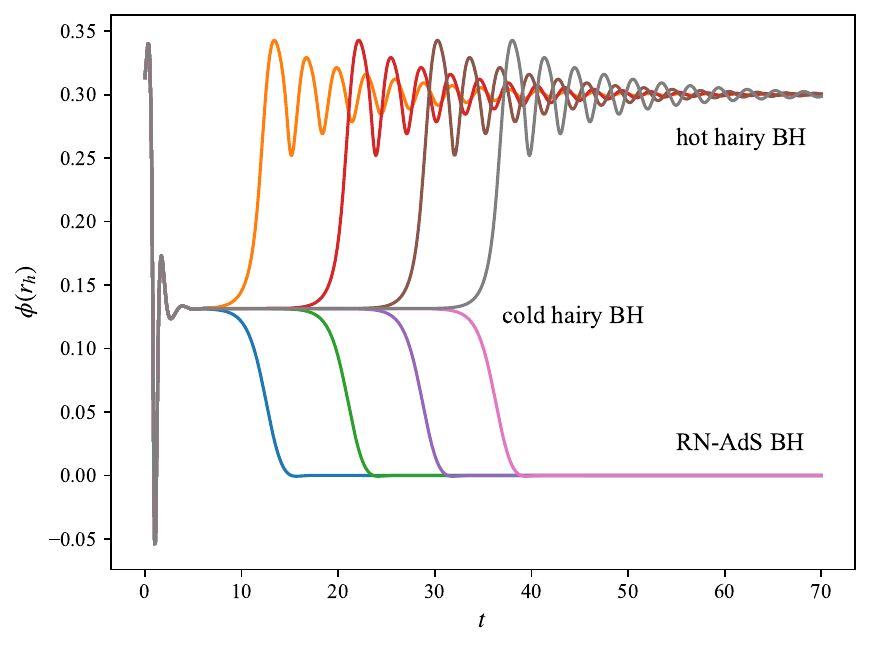}\label{fig:21}}
		\subfigure[]{\includegraphics[width=.49\linewidth]{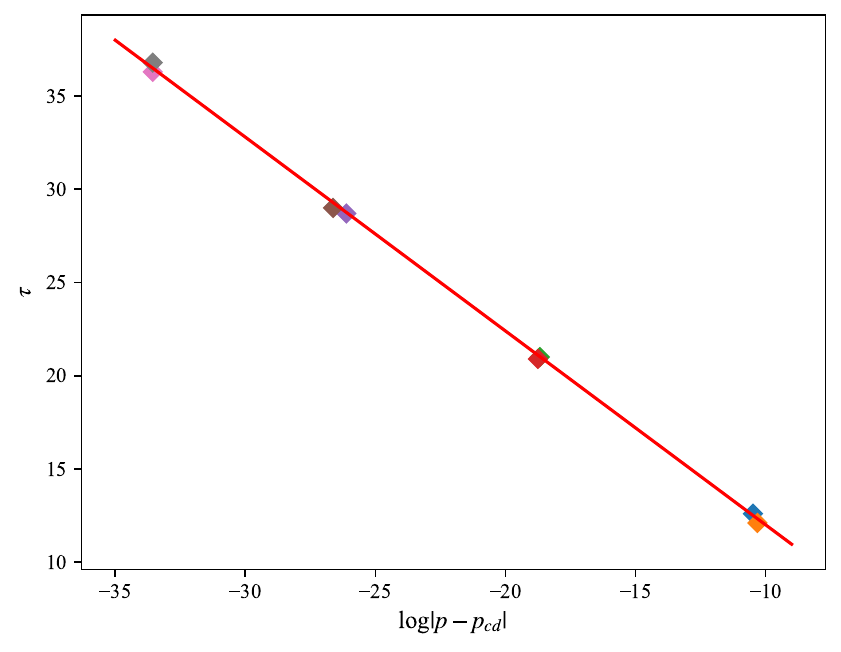}\label{fig:22}}
		\caption{(a): The value of the scalar field at the horizon as a function of time in the case where the accretion strength $p$ is close to the critical value $p_{cd}$.
				(d):  The time $\tau$ of the intermediate solution that stays near the cold hairy black hole with respect to ln$|p-p_{cd}|$.
				All the curves and points of the same color in the figure correspond to each other.}\label{fig:21-22}
	\end{center}
\end{figure}
%%%%%%%%%%%%%%%%%%

The evolution of the value of the scalar field at the apparent horizon is shown in figure \ref{fig:21}.
It turns out that there is a threshold $p_{cd}$ for the accretion strength to distinguish two different stable final states.
For the accretion process whose strength is less than the critical value, the final state of the evolution is still a hot hairy black hole but with more energy.
On the other hand, for the case where the accretion strength is greater than the critical value, a dynamical transition occurs, leading to the descalarization phenomenon.
That is to say, the accretion process of sufficient strength can also induce the excitation process from the ground state of the hot hairy black hole to the ground state of the RN-AdS black hole.
Near the threshold $p_{cd}$, similar to the critical scalarization phenomenon, the scalar field gradually converges to a static critical configuration after a short period of drastic changes.
This critical state is a cold hairy black hole with linear dynamical instability.
Similarly, the time that the critical state exists in the dynamical intermediate process depends on the difference between the accretion strength $p$ and the critical value $p_{cd}$.
The relationship (\ref{eq:5.2}) still holds, as shown in figure \ref{fig:22}.

%%%%%%%%%%%%%%%%%%
\begin{figure}
	\begin{center}
		\includegraphics[width=.5\linewidth]{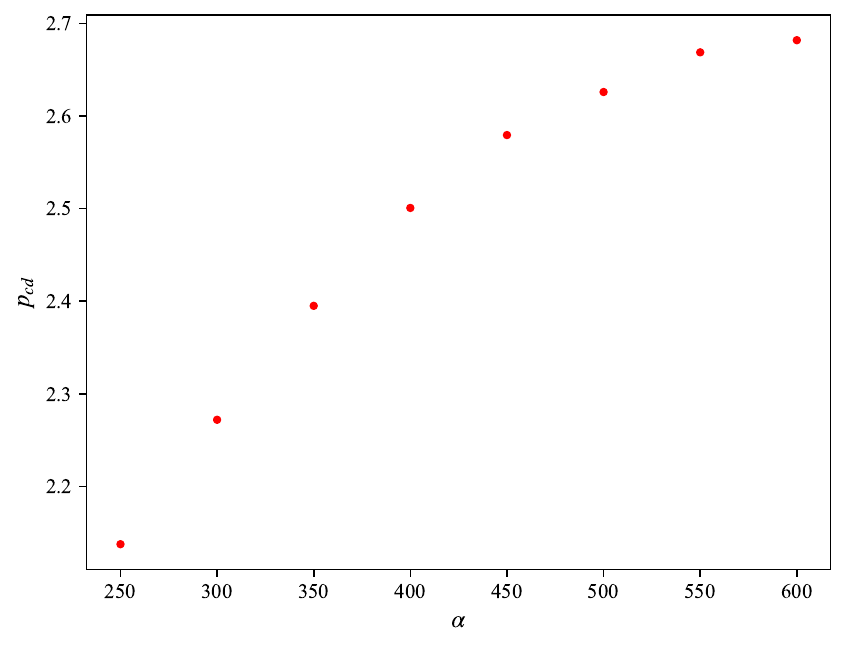}
		\caption{The critical value of accretion strength $p_{cd}$ as a function of the coupling constant $\alpha$.
			The energy density of the central hot hairy black hole is fixed to be $\epsilon=1.6$.}
		\label{fig:23}
	\end{center}
\end{figure}
%%%%%%%%%%%%%%%%%%

Different from the dynamical transition from a RN-AdS black hole to a hot hairy black hole, from the phase diagram in figure \ref{fig:3}, it can be seen that the entropy gap between the hot and cold hairy black holes increases with the coupling constant $\alpha$.
That is to say, the dynamical barrier for the transition from a hot hairy black hole to an RN-AdS black hole gradually increases with the coupling constant $\alpha$.
This is consistent with the numerical simulation results presented in figure \ref{fig:23}, which describe the monotonically increasing behavior of the accretion strength threshold with the coupling constant $\alpha$.

Through the above real-time dynamics, we realize the bidirectional dynamical transition process of a gravitational system between two local ground states.
By fine-tuning the parameter $p$ that characterizes the disturbance strength, the gravitational system will stay in a critical excited state with linear instability in the dynamical intermediate process, exhibiting critical dynamics.
In fact, such critical dynamics is universal and independent of the disturbance parameters.
That is to say, for the disturbance form described by (\ref{eq:5.3}), by fixing an appropriate value of parameter $p$, a similar critical dynamical process can also be triggered by fine-tuning the parameter $w$ or $z_{c}$.
Not only that, the critical dynamics is also universal to the disturbance form, as long as it can make the system cross the corresponding dynamical barrier.

\section{Dynamics of the excited state}\label{sec:BD}
In the previous section, we have revealed the real-time dynamics in the case where the initial central black hole is a linearly stable local ground state, where novel critical phenomena emerge.
In this section, we further investigate the dynamics in the case where the inital central black hole is an excited state with linear dynamical instability, in which the gravitational system exhibits a more interesting critical behavior.
For consistency, we take the cold hairy black hole in the ensemble with energy density $\epsilon=1.6$ as the initial configuration, in this case the hot hairy black hole with maximum entropy as the dominant thermal phase, and then impose the scalar field perturbation described by (\ref{eq:5.3}).

%%%%%%%%%%%%%%%%%%
\begin{figure}
	\begin{center}
		\subfigure[]{\includegraphics[width=.49\linewidth]{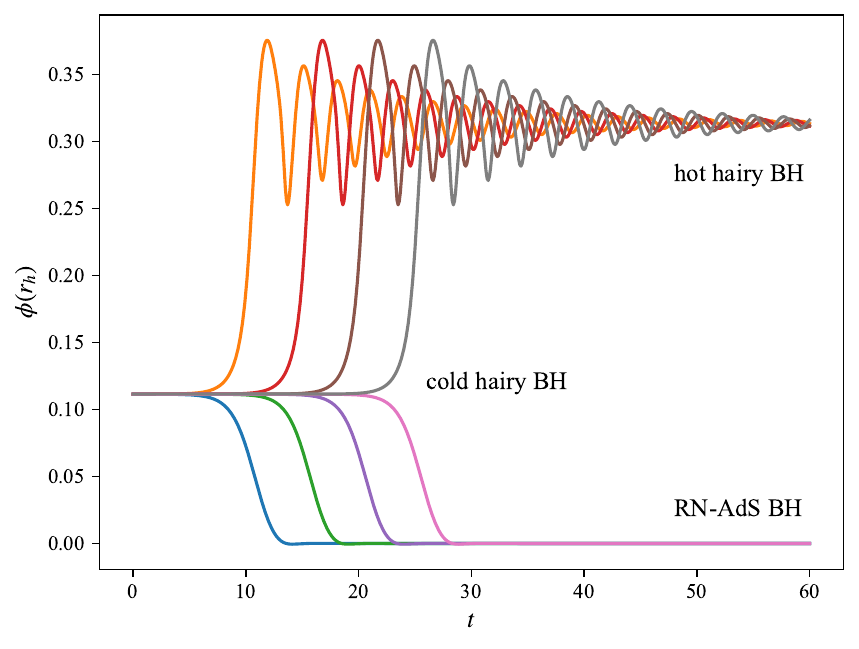}\label{fig:24}}
		\subfigure[]{\includegraphics[width=.49\linewidth]{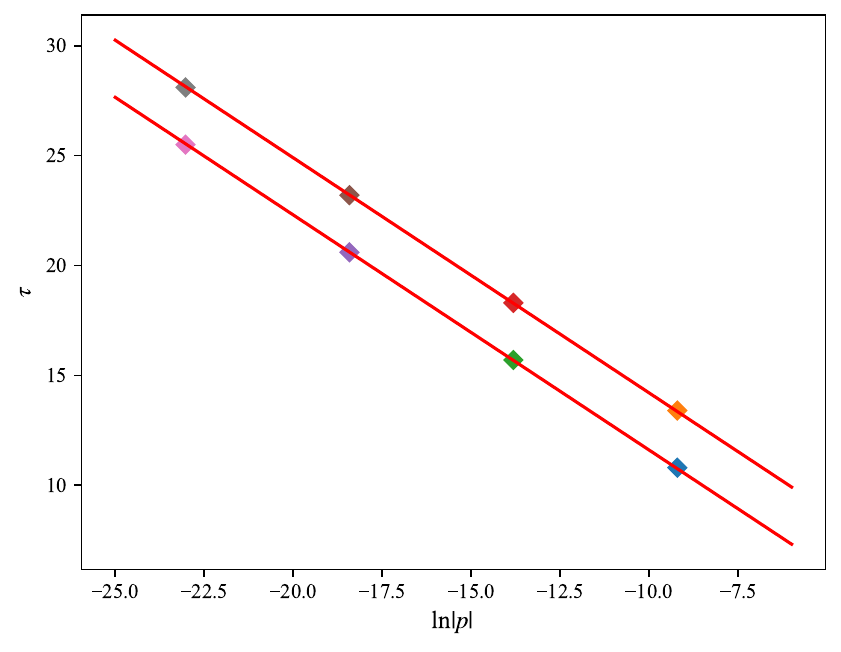}\label{fig:25}}
		\caption{(a): The value of the scalar field at the horizon as a function of time in the case where the accretion strength $p$ is close to zero.
			(d):  The time $\tau$ of the intermediate solution that stays near the cold hairy black hole with respect to ln$|p|$.
			All the curves and points of the same color in the figure correspond to each other.}\label{fig:24-25}
	\end{center}
\end{figure}
%%%%%%%%%%%%%%%%%%

Due to the linear instability of the initial gravitational system, the dynamical transition can occur under arbitrarily small perturbations.
Interestingly, there are two local ground states with linear dynamical stability that can serve as the final state of the dynamical evolution.
In the microcanonical ensemble, the entropy of the system describes the competitive relationship between thermal phases in equilibrium.
From a thermodynamic point of view, the system tends to reside in the state of maximum entropy, that is, the hot hairy black hole.
However, from the dynamics, it turns out that the system does not show a preference according to the entropy of the state.
From the numerical results in figure \ref{fig:24}, which show the evolution of the value of the scalar field at the apparent horizon, it can be seen that different values of the perturbation amplitude can still induce the system to evolve into two different stable final states.
We find that the final fate of the dynamical process corresponding to the positive perturbation amplitude is an RN-AdS black hole, on the contrary, the negative perturbation amplitude pushes the evolved system to a hot hairy black hole. 
Such a result indicates that the gravitational system undergoes a special class of critical dynamics with a perturbation strength threshold of zero $p_{*}=0$.
That is to say, in this case the critical state in the critical dynamical process is the initial state itself.
The corresponding dynamical barrier for the transition is zero.
Similarly, the smaller the perturbation strength, the longer the system will remain in the unstable initial state.
In this process, the relationship (\ref{eq:5.3}) holds, as shown in figure \ref{fig:25}.
Due to numerical errors, even with a perturbation amplitude that is strictly zero, a dynamical transition still occurs after long-term evolution, resulting in the values of $\tau$ corresponding to the scalarization and descalarization processes lying on two parallel lines, respectively.
Obviously, the slope of these two lines is equal to the reciprocal of the imaginary part of the single unstable mode of the initial cold hairy black hole.
The evolution of the dominant mode in the dynamical intermediate process is similar to that in figure \ref{fig:18}, with the difference being the lack of the first stage.

In order to exclude the influence of the thermodynamic potential of the local ground state on the conclusion, the real-time dynamics of the cold hairy black hole in the ensemble with energy density $\epsilon=2.2$, in which the RN-AdS black hole possesses the maximum entropy, is also studied.
Similar critical dynamical phenomena occur.
The unstable cold hairy black hole maintains its static configuration in the absence of perturbation, and evolves to a stable state through a non-equilibrium process with any small perturbation.
The specific configuration of the final state depends on the sign of the perturbation amplitude.
The selection of the final state is consistent with the case of energy density $\epsilon=1.6$: the plus sign corresponds to an RN-AdS black hole and the minus sign corresponds to a hot hairy black hole.
Such results indicate that the dynamical transition mechanism from an excited state to a local ground state is independent of the thermodynamic potential between the local ground states.

%%%%%%%%%%%%%%%%%%
\begin{figure}
	\begin{center}
		\subfigure[]{\includegraphics[width=.49\linewidth]{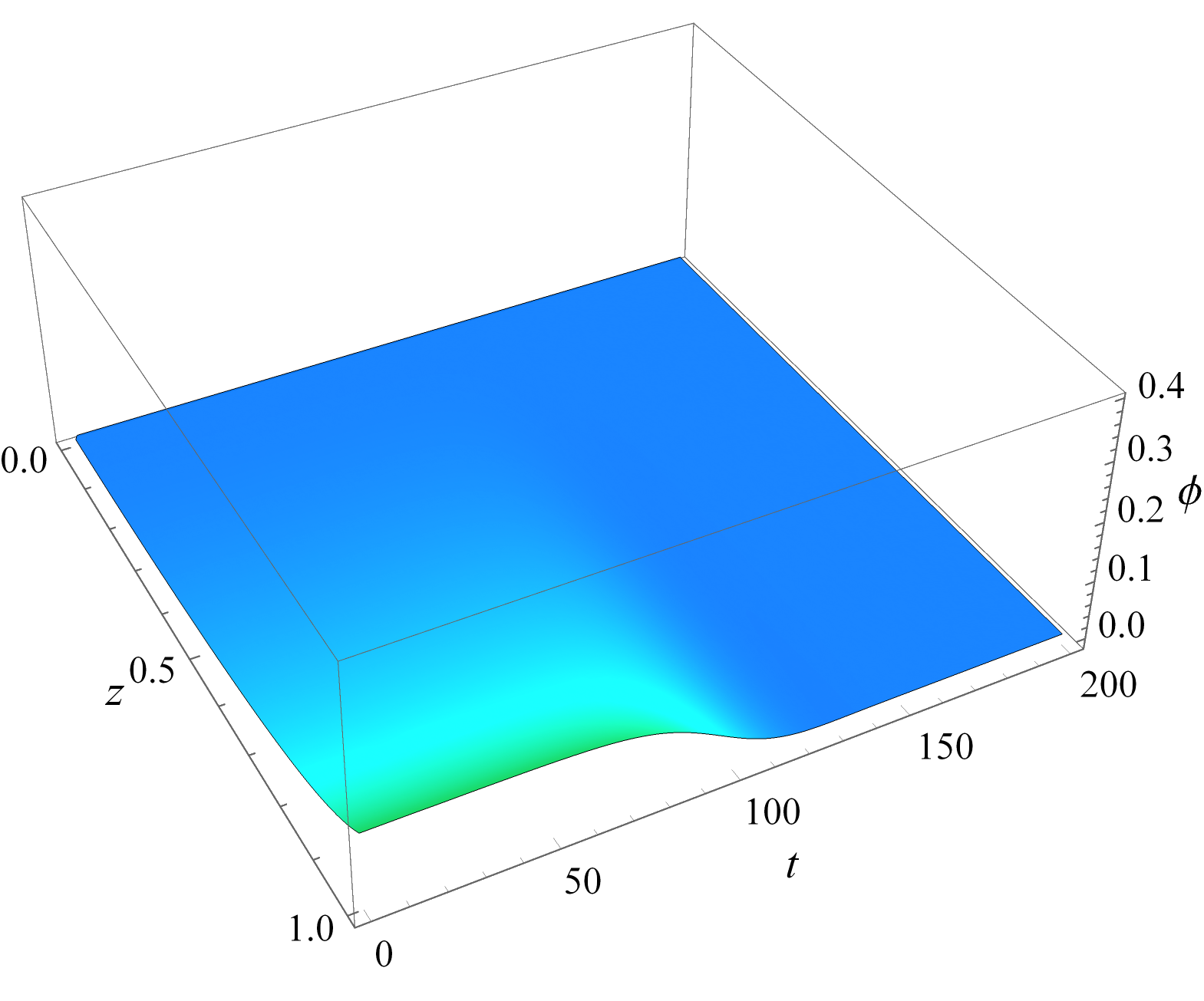}\label{fig:26}}
		\subfigure[]{\includegraphics[width=.49\linewidth]{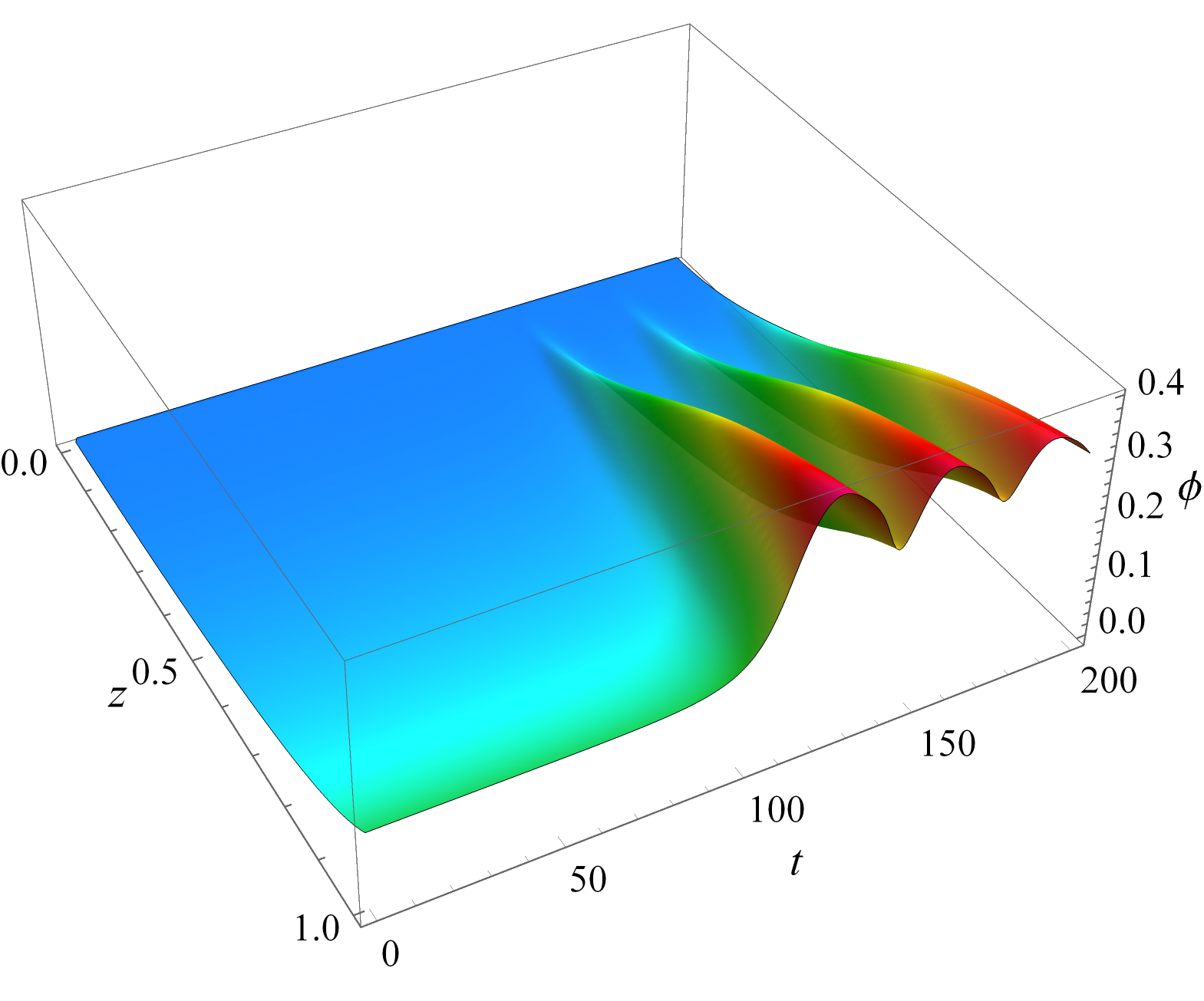}\label{fig:27}}
		\subfigure[]{\includegraphics[width=.49\linewidth]{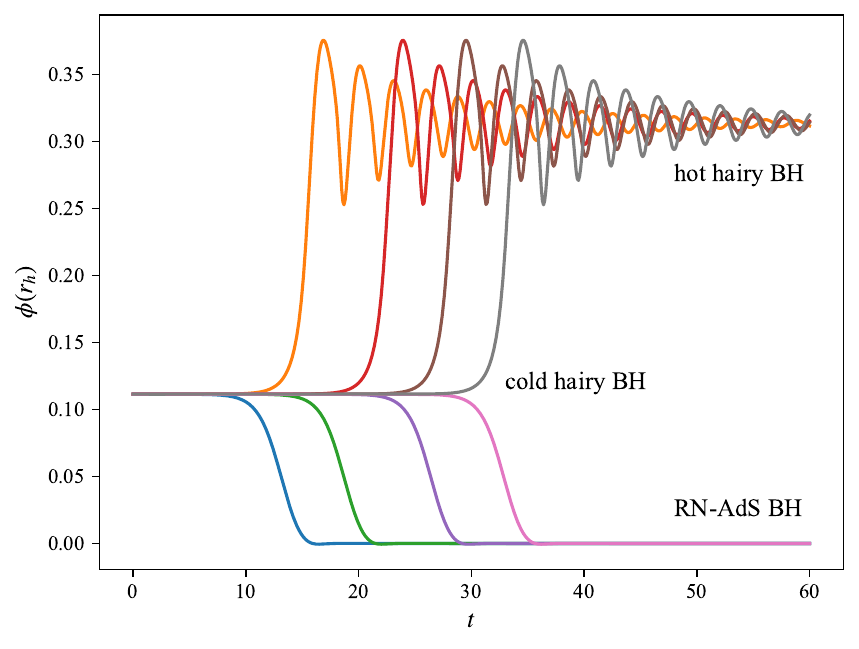}\label{fig:28}}
		\subfigure[]{\includegraphics[width=.49\linewidth]{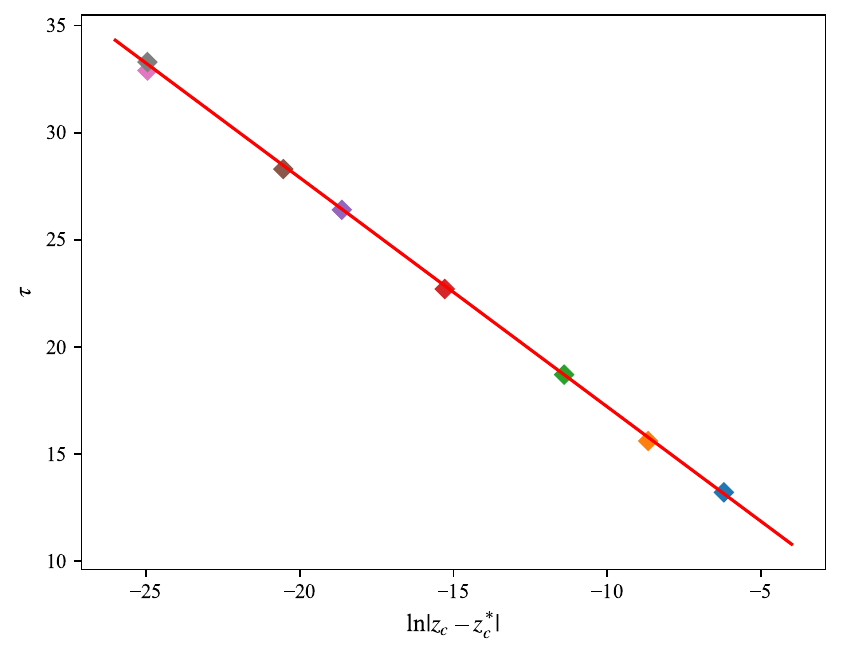}\label{fig:29}}
		\caption{(a, b): The profile of the scalar field as a function of time in the case of $z_{c}=0.9$ and $z_{c}=0.95$, respectively. 
				The perturbation amplitude is fixed as $p=+0.01$.
				(c): The value of the scalar field at the horizon as a function of time in the case where the parameter $z_{c}$ is close to the critical value $z^{*}_{c}$.
				(d):  The time $\tau$ of the intermediate solution that stays near the cold hairy black hole with respect to ln$|z_{c}-z^{*}_{c}|$.
				All the curves and points of the same color in the figure correspond to each other.}\label{fig:26-19}
	\end{center}
\end{figure}
%%%%%%%%%%%%%%%%%%

For the case where the initial configuration is unstable, a zero threshold of the perturbation strength for the critical dynamics is reasonable.
However, the relationship between the sign of the perturbation amplitude and the final state of evolution is somewhat puzzling.
In fact, we find that such a correspondence is related to the parameter $z_{c}$ in the perturbation (\ref{eq:5.3}), which characterizes the distance of the perturbation from the horizon.
For the cases of perturbations near the horizon and far away from the horizon, the corresponding relationship between the sign of the perturbation amplitude and the final state of the evolution is exactly opposite.
Fixing the perturbation amplitude $p=+0.01$ without loss of generality, we show the evolution of the scalar field configuration with time for the cases of $z_{c}=0.9$ and $z_{c}=0.95$ in figures \ref{fig:26} and \ref{fig:27}, respectively.
It turns out that a near-horizon perturbation with a positive sign leads to the formation of a hot hairy black hole instead of an RN-AdS black hole.
Such a result on the one hand shows that the selection of the final state is not only determined by the perturbation amplitude but also depends on the specific form of the perturbation, and on the other hand indicates that there should be a critical value for the parameter $z_{c}$ such that the gravitational system undergoes critical dynamics.
The critical value $z^{*}_{c}$, which is a function of perturbation amplitude $p$, is obtained by dichotomy, near which the system exhibits critical behaviors.
From the figure \ref{fig:28}, which shows the evolution of the value of the scalar field at the horizon, it can be seen that the time of the intermediate solution that stays near the critical black hole depends on the difference between the parameter $z_{c}$ and the critical value $z^{*}_{c}$.
Such a dependency satisfies (\ref{eq:5.2}), as shown in figure \ref{fig:29}.
Note that in this case the critical black hole that emerges in the dynamical intermediate process is not the one in figure \ref{fig:24}, which is exactly the initial state.
With the imposition of a perturbation of non-zero amplitude, the energy of the gravitational system changes even if the perturbation amplitude is so small, indicating that the perturbed system deviates from the original ensemble.
By fine-tuning the perturbation parameter $z_{c}$, the system converges to the cold hairy black hole in the new ensemble.
That is to say, such a perturbation is absorbed by the excited state without triggering its dynamical instability.
Certainly, the energy of the gravitational system is changed.
Since the dynamically unstable excited state can be viewed as a non-equilibrium configuration at a certain moment in the evolution process, the realization of the above process shows that the critical dynamics is quite universal to the initial value.
In fact, for any non-equilibrium configuration in a time-dependent process, the critical phenomena will also appear by applying appropriate perturbations.

%%%%%%%%%%%%%%%%%%
\begin{figure}
	\begin{center}
		\includegraphics[width=.5\linewidth]{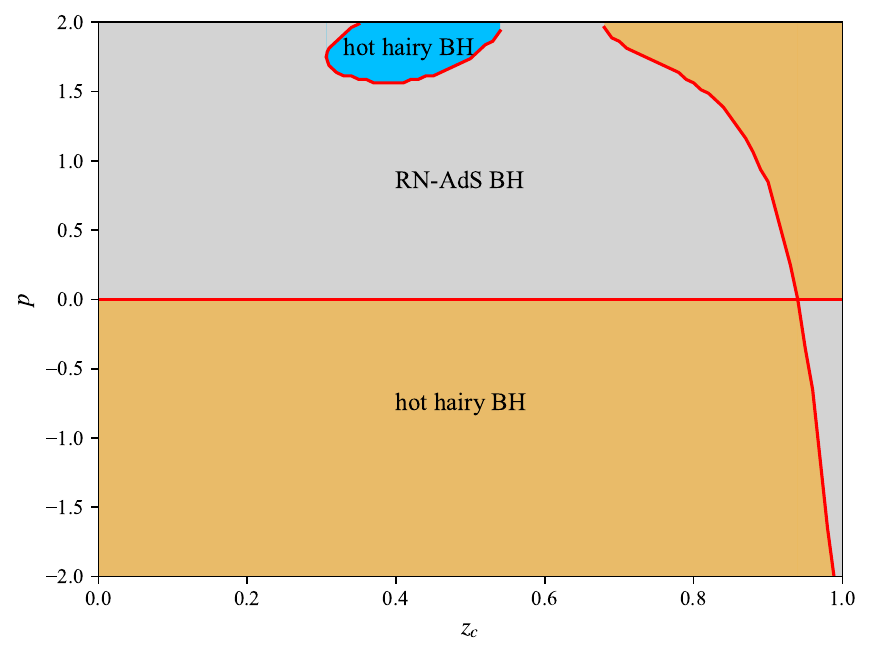}
		\caption{The final states of evolution in the parameter space $(z_{c}, p)$.
				Regions of different colors indicate different stable final states, among which the gray regions represent RN-AdS black holes, the orange regions represent hot hairy black holes with a positive scalar field, and the blue region represents hot hairy black holes with a negative scalar field.
				The domain walls between different regions correspond to unstable cold hairy black holes.}
		\label{fig:32}
	\end{center}
\end{figure}
%%%%%%%%%%%%%%%%%%

\begin{figure}
	\begin{center}
		\subfigure[]{\includegraphics[width=.49\linewidth]{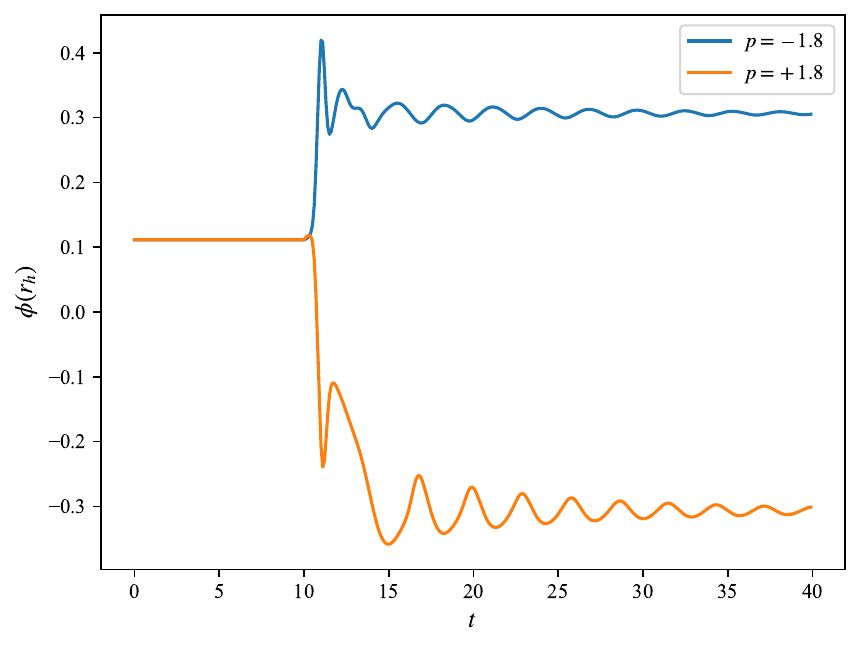}\label{fig:33}}
		\subfigure[]{\includegraphics[width=.49\linewidth]{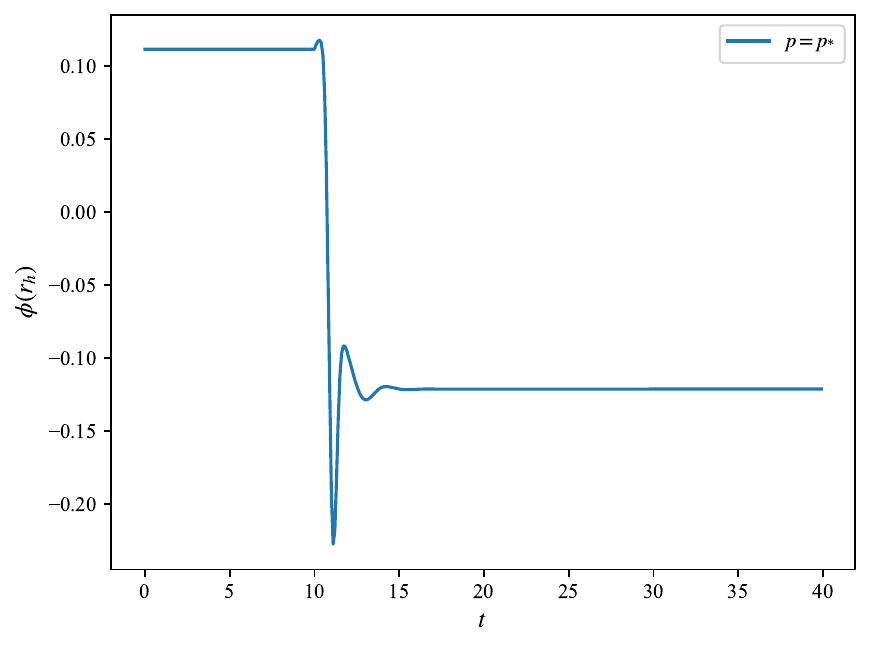}\label{fig:34}}
		\caption{The value of the scalar field at the horizon as a function of time.
				The perturbation with parameter $z_{c}=0.5$ is imposed at time $t=10$.
				The blue and orange lines in (a) represent the cases where the perturbation amplitude is $p=-1.8$ and $p=+1.8$ respectively, and the line in (b) corresponds to the critical value $p_{*}$ on the boundary of the blue region in figure \ref{fig:32}.}\label{fig:33-34}
	\end{center}
\end{figure}

The existence of the critical value $z^{*}_{c}$ indicates that there is another critical value other than zero for the perturbation amplitude $p$.
Since the combination of parameters $(z_{c}=z^{*}_{c},p=+0.01)$ corresponds to a critical black hole, naturally, fixed the parameter $z_{c}=z^{*}_{c}$, the gravitational system will also exhibit critical behaviors around $p=+0.01$.
In other words, $p=+0.01$ is also a threshold for the perturbation with parameter $z_{c}=z^{*}_{c}$.
In order to reveal the dependence of the final state of evolution on the parameters, we scan the parameter space $(z_{c}, p)$ and display the spectrum of the final state in figure \ref{fig:32}.
It turns out that the two stable local ground states occupy different regions of the parameter space, separated by the boundary representing the excited state.
If a one-parameter curve connects two different local ground states, there must be a point where it intersects the boundary.
This point is the threshold of critical dynamics.
Note that the points on the boundary $p=0$ represent the same state, namely the initial excited state.
The existence of other boundaries indicates that for an appropriate form of perturbation, there exists a series of specific combinations of parameters that fail to trigger the single unstable mode of the excited state.
Such a perturbation only changes the energy of the unstable gravitational system without changing its dynamical properties.
Interestingly, there is a region inhabited by the hot hairy black hole with a negative scalar field, whose boundary also corresponds to the cold hairy black hole with a negative scalar field.
Although there is degeneracy among the hot hairy black holes with positive and negative scalar fields due to the symmetry $\phi\rightarrow-\phi$, the dynamical processes from a cold hairy black hole with a positive scalar field to these two degenerate states are not the same, as shown in figure \ref{fig:33}.
For the perturbation parameters on the boundary of the blue region, the corresponding perturbations are absorbed by the original excited state with a positive scalar field and induce a dynamical transition to the other degenerate excited state with a negative scalar field, as shown in figure \ref{fig:34}. 

%%%%%%%%%%%%%%%%%%
%%%%%%%%%%%%%%%%%%

%=======================================================================

\section{Conclusion}\label{sec:Co}
In the EMs theory, the thermodynamic and dynamic properties of the gravitational system depend heavily on the interaction between the scalar and Maxwell fields.
For the case of the quadratic non-minimal coupling, the near-extremal RN-AdS black hole is dynamical unstable and can spontaneously scalarize to form a hairy black hole.
Since there is only one excited state and one global ground state, this dynamical transition process is unidirectional and dull.
Different from that, for the case of the non-minimal coupling function considered in our paper dominated by a quartic term, the richer phase structure leads to the emergence of many exciting dynamical processes.

In order to investigate the real-time dynamics, in the first step, we have revealed the phase structure of the model in the microcanonical ensemble.
The related results show that the domain of existence of solutions consists of a branch of RN-AdS black holes and two branches of hairy black holes, which are called as hot and cold hairy black holes, respectively.
Among them, the branch of cold hairy black holes is smoothly connected with the branch of RN-AdS black holes at the extremal RN-AdS black hole and has an upper limit of energy.
The branch of hot hairy black holes extends from this upper limit of energy to the over-extremal region.
For a gravitational system with fixed energy, the cold hairy black hole with the minimum entropy is in an excited state, and the RN-AdS black hole and hot hairy black hole are in two local ground states due to the larger entropy.

In the second step, we have studied the effective potentials and the quasinormal modes of these three classes of thermal phases.
For both local ground states, the effective potential exhibits a potential barrier near the event horizon.
However, for the excited state, there is a negative region in the effective potential near the event horizon.
Such a potential well is generally associated with the tachyonic instability.
From the linear perturbation theory, only the excited state possesses an unstable mode with an imaginary part greater than zero, indicating the dynamical instability.
Both local ground states are dynamically stable at the linear level.

In the third step, the real-time dynamics based on the two local ground states are revealed.
By simulating the fully nonlinear accretion process of a scalar field to a central black hole, we have discovered that the gravitational system can dynamically transition between the two local ground states.
Moreover, there is a dynamical barrier in such a transition process, which is reflected in the existence of a threshold for the accretion strength $p$.
For the case of the accretion strength less than the threshold, the scalar field disturbance is absorbed by the central black hole, increasing the energy without changing the essential properties.
On the other hand, the accretion process with the strength greater than the threshold induces a drastic change in the gravitational configuration and triggers the corresponding transition process.
Near the threshold, the gravitational system is attracted to an excited state in the dynamical intermediate process, and the time to maintain it increases continuously as the accretion strength approaches the threshold.
Interestingly, the dynamical barrier that needs to be overcome to trigger the transition process of RN-AdS black holes decreases with the increase of the coupling strength between the scalar and Maxwell fields, which is just the opposite of the case of hot hairy black holes.

In the final step, we have investigated the real-time dynamics with the excited state as the initial value.
On the one hand, due to the linear dynamical instability, there exists a special kind of critical dynamics with a zero threshold for the perturbation strength. 
The perturbation amplitudes with different signs push the gravitational system to fall into different stable local ground states.
The specific selection of the final state of evolution depends on other parameters of the perturbation.
On the other hand, for the perturbation with a fixed non-zero amplitude, the parameter $z_{c}$ describing the position of the perturbation can also induce the critical phenomenon.
Such a result indicates that the perturbation with the threshold parameters fails to trigger the single unstable mode of the corresponding excited state, but only changes its energy after being absorbed by the central black hole.
Further the spectrum of the final state of evolution is revealed in the parameter space $(z_{c}, p)$.
The two linearly stable local ground states occupy different regions respectively, bounded by the dynamically unstable excited state.

The research on the dynamics of gravitational systems has formed a standard framework, from the thermodynamic properties of the equilibrium state, to the linear stability analysis of the near-equilibrium state, all the way to the real-time dynamics simulation of the far-from-equilibrium state.
This paper demonstrates this procedure by taking an EMs gravitational system in AdS spacetime as an example.
Different from the spontaneous process of the unstable system, the excitation process of the stable ground state shown in this paper has more practical significance and observable effects due to the stability of objects in the real world.
Although the EMs model is unlikely to be relevant to astrophysics, it has important applications to holography in the AdS spacetime.
At present, some related dynamical studies have shown that such critical phenomena with dynamical barrires widely exist in various gravitational systems, such as gravitational collapse \cite{Choptuik:1992jv,Abrahams:1993wa,Evans:1994pj,Koike:1995jm,Gundlach:1995kd,Liebling:1996dx,Choptuik:1996yg,Brady:1997fj,Bizon:1998kq,Garfinkle:1998va,Choptuik:2004ha,Gundlach:2007gc}, EMs \cite{Zhang:2021nnn,Zhang:2022cmu}, EsGB \cite{Liu:2022fxy} and holographic first-order phase transition \cite{Li:2020ayr,Bea:2020ees,Chen:2022cwi} models.
In addition, one can observe that the phase structure here has certain resemblances to those between the vacuum black rings and Myers-Perry black holes in higher-dimensional spacetimes \cite{Myers:1986un,Emparan:2001wn,Emparan:2006mm,Emparan:2007wm,Emparan:2008eg}, thus it is expected that there will be similar critical dynamics phenomena.
On the other hand, for a specific self-interaction potential of a scalar field, there are indications of the existence of multiple local ground states \cite{Herdeiro:2020xmb,Hong:2020miv}, leading to the emergence of critical dynamics.
For astronomical observations, the case of a rotating black hole with non-minimal coupling to gravity is the most favorable candidate \cite{Cunha:2019dwb,Dima:2020yac,Herdeiro:2020wei,Berti:2020kgk}, and the dynamical behaviors during the corresponding critical transition process can be characterized by gravitational waves.

\appendix

\section*{Acknowledgement}
This research is partly supported by the Natural Science Foundation of China (NNSFC)
under Grant Nos. 11975235, 12005077, 12035016, 12075202 and Guangdong Basic and Applied Basic Research
Foundation under Grant No. 2021A1515012374.
%=======================================================================
\bibliography{references}

\end{document}